\documentclass[12pt,twoside]{article}
\usepackage[dvips]{graphicx}
\usepackage{float}
\usepackage{latexsym}
\catcode`\@=11
\def\newsymbol#1#2#3#4#5{\let\next@\relax%
 \ifnum#2=\@ne\else%
 \ifnum#2=\tw@\let\next@\msyfam@\fi\fi%
 \mathchardef#1="#3\next@#4#5}
\def\mathhexbox@#1#2#3{\relax%
 \ifmmode\mathpalette{} {\m@th\mnnathchar"#1#2#3}
 \else\leavevmode\hbox{$\m@th\mathchar"#1#2#3$}\fi}
\font\tenmsy=msbm10
\font\sevenmsy=msbm7
\font\fivemsy=msbm5
\newfam\msyfam
\textfont\msyfam=\tenmsy
\scriptfont\msyfam=\sevenmsy
\scriptscriptfont\msyfam=\fivemsy
\edef\msyfam@{\hexnumber@\msyfam}
\def\Bbb#1{\fam\msyfam\relax#1}
\catcode`\@=\active

\newtheorem{theorem}{Theorem}[section]
\newtheorem{proposition}[theorem]{Proposition}
\newtheorem{lemma}[theorem]{Lemma}
\newtheorem{corollary}[theorem]{Corollary}

\newtheorem{example}[theorem]{Example}
\newtheorem{remark}[theorem]{Remark}

\newcommand{\ZZ}{{\cal K}_1^{[0,2t]}(0)}
\newcommand{\kkx}{{\cal K}_1^{[0,t]}(x)}
\newcommand{\ZK}{{\cal K}_0^{[0,t]}(0)}
\newcommand{\kkz}{\kk{0,t}}
\newcommand{\eq}[1]{\begin{equation}\label{#1}}
\newcommand{\en}{\end{equation}}
\newcommand{\eqn}{\begin{eqnarray*}}
\newcommand{\enn}{\end{eqnarray*}}
\newcommand{\eqnn}{\begin{eqnarray}}
\newcommand{\ennn}{\end{eqnarray}}
\newcommand{\proof}{{\noindent \it Proof:\ }}
\newcommand{\qed}{\hfill {\bf qed}\par\medskip}
\newcommand{\BR}{{{\Bbb R}^d }}
\newcommand{\bi}{\begin{description}}
\newcommand{\ei}{\end{description} }
\newcommand{\CC}{{{\Bbb C}}}
\newcommand{\RR}{{\Bbb R}}
\newcommand{\TPP}[1]{\lk e^{-\frac{#1}{n} V} e^{-\frac{#1}{n}K_0}e^{-\frac{#1}{n}\hf}\rk^n}
\newcommand{\T}[1]{e^{- #1 K}} 
\newcommand{\TT}[1]{e^{-#1 \tilde K}}

\newcommand{\llk}{{\ell_k}}
\newcommand{\llx}{\ell_x}

\newcommand{\can}{(\Phi_0, \prod_{j=1}^m O_j(P_{j-1}) \Phi_j)_\fffb}
\newcommand{\bl}[1]{\begin{lemma}\label{#1}}
\newcommand{\el}{\end{lemma}}
\newcommand{\bc}[1]{\begin{corollary}\label{#1}}
\newcommand{\ec}{\end{corollary}}
\newcommand{\bt}[1]{\begin{theorem}\label{#1}}
\newcommand{\et}{\end{theorem}}
\newcommand{\bp}[1]{\begin{proposition}\label{#1}}
\newcommand{\ep}{\end{proposition}}
\newcommand{\br}[1]{\begin{remark}\label{#1}}
\newcommand{\er}{\end{remark}}

\newcommand{\limn}{\lim_{n\rightarrow\infty}}

\newcommand{\limt}{\lim_{t\rightarrow\infty}}

\newcommand{\kak}[1]{(\ref{#1})}
\newcommand{\LR}{{L^2(\BR)}}
\newcommand{\K}[1]{[#1]}
\newcommand{\LRT}{{L^2(\RR^{d +1})}}
\newcommand{\LRTT}{{L^2(\RR^{d +2})}}

\newcommand{\fff}{{{\cal F}_{\rm b}}}
\newcommand{\fffb}{{{\cal F}_{\rm b}}}
\newcommand{\fffi}{{\cal F}_{\rm b}^\infty}
\newcommand{\ffff}{{\cal F}_{\rm b, fin}}
\newcommand{\is}{\inf\sigma}
\newcommand{\ehtp}{{e^{-t\fri(P)}}}
\newcommand{\f}{^{-1}}
\newcommand{\lk}{\left(}
\newcommand{\rk}{\right)}
\newcommand{\lkk}{\left\{}
\newcommand{\rkk}{\right\}}

\newcommand{\add}{a^{\ast}}

\newcommand{\pb}{e^{iP\cdot b(t)}}
\newcommand{\ppf}{e^{-i\pf \cdot b(t)}}

\newcommand{\ass}{a^\sharp}

\newcommand{\ov}[1]{\overline{#1}}
\newcommand{\hf}{H_{\rm f}}
\newcommand{\pf}{{P_{\rm f}}}

\newcommand{\gr}{\varphi_{\rm g}}

\newcommand{\half}{\frac{1}{2}}
\newcommand{\han}{{1/2}}

\newcommand{\dg}[1]{d\Gamma(#1)}
\newcommand{\dgg}[1]{\Gamma_{\#}(#1)}
\newcommand{\dggg}[1]{d\Gamma_{\#}(#1)}
\newcommand{\DG}[1]{d\Gamma_{0}(#1)}

\newcommand{\vvv}[1]{\lk\!\!\!\begin{array}{c}#1\end{array}\!\!\!\rk}
\newcommand{\MMM}[4]
{\lk\!\!\!\begin{array}{cc}#1&#2\\ #3&#4\end{array}\!\!\!\rk}

\newcommand{\MMMM}[9]
{\lk\!\!\!\begin{array}{ccc}#1&#2&#3\\ #4&#5&#6\\#7&#8&#9\end{array}\!\!\!\rk}

\newcommand{\la}{{\tilde\varphi}}
\newcommand{\s}{\sigma}
\newcommand{\hhh}{{\cal H}}
\newcommand{\xxx}[2]{
e^{-(s_{#1}-s_{#2})K} e^{-(t_{#1}-t_{#2})H(P_{#2})}
}
\newcommand{\yyy}[1]{e^{-i\xi_{#1}\cdot \tot}}
\newcommand{\yyyy}[1]{e^{-i\xi_{#1} P_{#1}}}
\newcommand{\xxxx}[2]{
e^{-(s_{#1}-s_{#2})K} e^{-(t_{#1}-t_{#2})\fri(P)}}

\newcommand{\jjj}{\sum_{j=1}^{d -1}}

\renewcommand{\d}{\displaystyle}
\newcommand{\av}{{A_\vp}}
\newcommand{\vp}{{\hat \varphi}}

\newcommand{\sso}{\stackrel{d }{\oplus}}
\newcommand{\dep}{\delta^\perp}
\newcommand{\non}{\nonumber}
\newcommand{\cd}{{\cal D}}

\newcommand{\fri}{{H}}
\newcommand{\ffri}{{H_{\rm F}}}
\newcommand{\QPP}{{Q\otimes 1-1\otimes \pf}}
\newcommand{\tot}{{P_{\rm T}}} 
\newcommand{\Hs}{{H_{\rm S}}}
\newcommand{\hs}{{H_\sigma}}
\newcommand{\z}{{\Bbb Z}_\han}
\newcommand{\kk}[1]{{\cal K}_1^{[#1]}(0)}
\newcommand{\kko}[1]{{\cal K}_0^{[#1]}(0)}
\renewcommand{\t}{\vartheta}
\newcommand{\tf}{\vartheta^{-1}}

\makeatletter
\@addtoreset{equation}{section}
\makeatother

\title
{Fiber Hamiltonians in the non-relativistic quantum electrodynamics}

\author{
Fumio Hiroshima\thanks{e-mail: hiroshima@ math.kyushu-u.ac.jp}
\\ 
Faculty of Mathematics,
Kyushu University, \\
Hakozaki, Higashi-ku, 6-10-1, \\
 Fukuoka 812-8581, Japan. 
}
\date{\today}
\begin{document}
\pagestyle{myheadings}
\markboth{Fiber Hamiltonians}{Fiber Hamiltonians}

\setlength{\baselineskip}{15pt}
\maketitle

\begin{abstract}
A translation invariant Hamiltonian $H$ in the nonrelativistic quantum electrodynamics is studied.
This Hamiltonian is decomposed with respect to the total momentum $\tot$:
$$H=\int_{\BR} ^\oplus \fri(P) dP,$$
where the self-adjoint fiber Hamiltonian $\fri(P)$ is defined 
for arbitrary values of coupling constants. 
It is discussed a relationship between rotation invariance of $H(P)$ and polarization vectors, 
and functional integral representations of
$n$ point Euclidean Green functions of $H(P)$  
is given. 
From these,  some applications concerning with degeneracy of ground states,  ground state energy and  
expectation values of suitable observables with respect to ground states are given. 
\end{abstract}

{\footnotesize 
\tableofcontents
}

\section{Introduction and statements of results}
In this paper we investigate self-adjoint operator $\fri(P)$ indexed by $P\in\BR$ 
by means of a functional integral representation of $e^{-t\fri(P)}$. 
Operator $H(P)$ is 
derived from a translation invariant self-adjoint operator $H$ acting in a Hilbert space
$\hhh$ such  that 
\eq{tr}
[H, \tot_j]=0,\quad j=1,...,d,\en
where $\tot=(\tot_1,...,\tot_d)$ denotes the $d$-tuple of the total momentum operators  
with   
$\s(\tot_j)=\RR$. Here  $\s(T)$ denotes the spectrum of $T$.  
Hence $\hhh$ and $H$ can be represented by constant fiber direct integrals: 
\eq{dd1}
\hhh\cong \int^\oplus_\BR \hhh(P) dP,\quad 
H\cong \int^\oplus_{{\Bbb R}^d} \fri(P) dP.
\en 
In this  paper we study the so-called Pauli-Fierz model  
in the nonrelativistic quantum electrodynamics, which describes 
an interaction between 
a quantum mechanical particle (electron) and a quantized radiation field. 
The Hamiltonian $H$ of this system is  defined 
as a self-adjoint operator minimally coupled to the 
quantized radiation field, which acts  in 
$\hhh:=\LR\otimes \fffb$, 
where $\fffb$ is a boson Fock space.  
We impose an ultraviolet cutoff on $H$ and 
work in the Coulomb gauge with 
$d-1$ polarization vectors. 
It is seen that $H$ without external potentials is translation invariant. Namely 
$H$ satisfies \kak{tr} for some total momentum operators,  
from which  $\hhh$ and $H$ can be decomposed such as 
\kak{dd1}. 
We shall show  that 
$\hhh(P)$ is unitarily equivalent to $\fffb$ and 
$\fri(P)$ is realized as a self-adjoint 
operator acting in $\fffb$ for each $P\in\BR$.

\subsection{Statements of results}
In this article we shall investigate  
(1) functional integral representations, 
(2) self-adjointness and essential self-adjointness, 
(3) ergodic properties of $e^{-tH(0)}$,  
(4) rotation invariance,
(5) energy inequalities and 
(6) measures associated with  ground states. 
(2),(3),(5) and (6) are studied through the functional integration (1). 
 
{\bf (1) Functional integrations:} \\
The polaron model $H_{\rm polaron}(P)$  is a typical example of fiber Hamiltonians, which is studied 
in \cite{sp5, sp2}  by functional integrals. 
In \cite[Appendix]{sp5} the functional integral representation of 
$(\Omega, e^{-tH_{\rm polaron}(P)}\Omega)$ is shown, where $\Omega$ denotes a vacuum vector in a boson Fock space. 
Our motivation to construct  \kak{st} below comes from this. 
In \cite{ffg, fp, h4} a functional integral representation of 
$(F, e^{-tH}G)_{\hhh}$ is given. 
It is the main achievement in this paper that a functional integral representation of 
$(\Psi, e^{-t\fri(P)}\Phi)_\fffb$ is 
constructed for arbitrary total momentum $P\in\BR$ and arbitrary values of coupling constants  
on a probability space $W\times Q_1$ equipped with 
the product measure, $db \otimes d\mu_1$, where $db$ is a measure associated with the 
particle and $d\mu_1$ with the quantized radiation field.
Although for example, it can be taken $C([0,\infty);\BR)$ as $W$ and 
the direct sum of the set of real Schwarz  distributions, 
$\oplus^d{\cal S}_{\rm real}'(\RR^{d+1})$, 
as $Q_1$, we do not specify them in this paper.  
See e.g., \cite{ffg, h4} for a detail. 
Moreover a functional integral representation of  
an $n$ point Euclidean Green function of the 
form  
\eq{FFFFF}
(\Phi_0, \prod_{j=1}^n \xxx j {j-1}  \Phi_j)_\fffb
\en
is also given. 
Here $K$ denotes a second quantized operator and $\Phi_j$, $j=1,...,n-1$, 
bounded multiplication operators.  
Since the interaction of the Pauli-Fierz Hamiltonian is introduced as 
a minimal coupling, 
we need a Hilbert space-valued stochastic integral 
to construct  the functional integral representation of 
$(\Psi, e^{-t\fri(P)}\Phi)_\fffb$.  
Actually  we show that 
\eq{st}
(\Psi, e^{-tH(P)}\Phi)_\fffb 
=\int_{W\times Q_1} \ov{\Psi_0}\Phi_t 
e^{-i\int_0^t{\cal A}_1 db(s)}e^{iP\cdot b(t)}db\otimes d\mu_1,
\en 
where the right-hand side above is in the Schr\"odinger representation instead of  the Fock representation, 
$(b(s))_{0\leq s}$ the $d$ dimensional Brownian motion with respect to $db$ 
and $\Psi_0$, $\Phi_t$ denotes some vectors.  
The integral  $\d \int_0^t {\cal A}_1 db(s)=\sum_{\mu=1}^d\int_0^t {\cal A}_{\mu,s} db_\mu(s)
$
denotes 
a Hilbert space-valued stochastic integral.   
See Section 3 for details. 
As far as we know 
it is the first time to give   functional integral 
representations explicitly such as \kak{st} of a fiber Hamiltonian minimally coupled to 
a quantized radiation field.

{\bf (2) Self-adjointness and essential self-adjointness:}\\
 In \cite {h11, h16}, applying a functional integral representation, 
we established the self-adjointness of $H$ for arbitrary values of coupling constants. 
The self-adjointness of $H(P)$ follows from that of $H$, which was done in \cite{lms}. 
In this paper as an application of the functional integral representation,
we show  the essential self-adjointness of a more singular operator $K(P)$, $P\in\BR$,  which is defined as $H(P)$ without 
the free field Hamiltonian $\hf$, is shown in Theorem \ref{poo}. 
The idea is to find an invariant domain by using \kak{st}.

{\bf (3) Ergodic properties and the uniqueness of ground states:} \\
The multiplicity of the ground state of the Pauli-Fierz Hamiltonian  
is estimated in e.g., \cite{h1, hisp2}. 
In \cite{hisp2} for a sufficiently small coupling constant, 
the uniqueness of the ground state of $H(P)$ is proved. We want to extend it for an arbitrary values of coupling constants. 
One merit of working in 
the Schr\"odinger representation is to define the positive cone: 
$${\cal K}_+=\{\Psi\in\hhh| \Psi\geq 0\},\quad 
{\cal K}_+^0=\{\Psi\in\hhh |\Psi>0\}\subset{\cal K}_+.$$
We say that bounded operator $T$ is positivity preserving if and only if 
$T{\cal K}_+\subset {\cal K}_+$ and positivity improving if and only if 
$T[{\cal K}_+\setminus\{0\}]\subset{\cal K}_+^0$. 
We discuss some positivity properties of $e^{-tH(0)}$. 
One important translation invariant model is the so-called Nelson model 
$H_{\rm N}(P)$ \cite{ne3} acting in 
a Hilbert space $\hhh_N$ with a fixed total momentum $P\in\BR$.   
In a simpler way than  \kak{st} 
we can also give 
the functional integral representation of $e^{-tH_{\rm N}(P)}$. 
Since the interaction term of the Nelson model   is linear,  
the integrand of the functional integral representation 
of $(\Psi, e^{-tH_{\rm N}(P)}\Phi)_{\hhh_N}$ is given 
by a  Riemann integral instead of the stochastic integral as in \kak{st} with the form 
\eq{ri}
(\Psi, e^{-tH_{\rm N}(P)}\Phi)_{\hhh_N}
=\int_{W\times Q_N} \ov{\Psi_0}\Phi_t 
e^{-\int_0^t\phi  ds}e^{iP\cdot b(t)} 
db\otimes d\mu_N,
\en 
where $(Q_N, d\mu_N)$ is some probability space. 
In \cite{gr1}, by a positivity preserving argument 
and a hypercontractivity argument, 
the uniqueness of ground state of $H_{\rm N}(0)$ is proven. 
See also \cite{sl, bo}. 
Since $e^{-\int_0^t\phi ds}e^{iP\cdot b(t)}$ is strictly positive for $P=0$ 
in the Schr\"odinger representation, 
it can be shown  that,  by \kak{ri},  $e^{-tH_{\rm N}(0)}$ is positivity improving.    
From this it can be also concluded that the ground state of $H_{\rm N}(0)$ is unique 
due to the infinite dimensional version 
of the Perron Frobenius theorem.  

Although we want to apply the Perron Frobenius theorem to  
the Pauli-Fierz model, 
we can not apply directly the positivity argument as for  the Nelson model 
\kak{ri},  
since $e^{-i\int_0^t{\cal A}_1 db(s)}$ in \kak{st} can not be only positive but also  real. 
Let us consider the multiplication operator $T_t=e^{itx}$, $t\in\RR$, 
in $L^2(\RR_x)$.  Although $T_t$ is not positivity preserving operator, 
${\Bbb F}T_t{\Bbb F}\f$, 
where 
${\Bbb F} $ denotes the Fourier transformation on $L^2(\RR)$, 
turns out to be a shift operator, i.e., 
$(f, {\Bbb F}T_t{\Bbb F}\f g)_{L^2(\RR)}=(f, g(\cdot+t))_{L^2(\RR)}\geq 0
$ 
for nonnegative functions 
$f$ and $g$. 
Then 
${\Bbb F}T_t{\Bbb F}\f$ is a positivity preserving operator, 
but {\it not} a positivity improving operator:
$$\begin{array}{ccc}
T_t=e^{itx}&\longrightarrow&{\Bbb F} T_t {\Bbb F}\f\\
{\rm multiplication} &\ \  & {\rm shift}
\end{array}
$$
 
This idea was applied to $H$ in \cite{h9}. 
In this paper we also do this for $H(0)$.  
We can show that $\t e^{-i\int_0^t{\cal A}_1 db(s)}\tf$ is  positivity preserving 
for some unitary operator $\t$ discussed in 
\cite{hirokawa, h9},   
which corresponds to 
the Fourier transformation on $\fffb$.    
Actually $\t=\exp(i(\pi/2)N)$, where $N$ denotes the number operator. 
Hence we can see that  by the functional integral representation \kak{st}, 
$\t e^{-tH(0)}\tf$  is 
a positivity {\it improving} operator in Theorem \ref{m2}, i.e., 
$$\t e^{-tH(0)}\tf[{\cal K}_+\setminus\{0\}]\subset{\cal K}_+^0.$$
As a corollary, 
the uniqueness of the ground state of $H(0)$ is shown 
for arbitrary values of coupling constants if it exists.

{\bf (4) Rotation invariance and the degeneracy of ground states:}\\
Operator $H(P)$ has also a rotational symmetry.
When the Hamiltonian includes a spin, a lower bound of the multiplicity, $M$, 
 of ground states can be  estimated 
by using this  rotational symmetry.  
Let us add a spin to $H$ which is denoted by 
$\hs$. 
It can be  shown that $\hs$ with  suitable polarization vectors 
is rotation invariant around some unit vector   $n\in\RR^3$, 
which is inherited to operator $\hs(P)$ with fixed total momentum $P$ 
acting in ${\Bbb C}^2\otimes\fffb$. Then we shall see that 
$\hs(P)$  is also decomposed with respect to the spectrum of the generator 
of the rotation around $n$, namely   
\eq{1212}
\hs(P)\cong \hs(|P|n)=\bigoplus_{z\in{\Bbb Z}_\han} \hs(P,z),\ \ \ {\Bbb C}^2\otimes \fffb=\bigoplus_{z\in{\Bbb Z}_\han}\fffb(z),
\en 
where $\cong$ denotes an unitary equivalence and ${\Bbb Z}_\han$  the set 
of 
half integers.   
Although  for a sufficiently small coupling constant, $M\geq 2$ is established in \cite{hisp2}, 
applying the decomposition \kak{1212} and \cite{sasa}, we can see that  $M\geq 2$ for  arbitrary values of coupling constants. 
See Corollary \ref{sasa2}.

{\bf (5) Energy inequalities:} \\
As is seen in \kak{st}, $P$ dependence on the integrand 
is just the exponent of the phase: $e^{iP\cdot b(t)}$. 
Trivial bound  $|e^{iP\cdot b(t)}|\leq 1$ and
 $|\t e^{-i\int_0^t{\cal A}_1db(s)}\tf \Psi|\leq \t e^{-i\int_0^t{\cal A}_1db(s)}\tf |\Psi|$ 
are useful to 
estimate the ground state energy of $H(P)$ from below. Then it can be shown that 
$\is(H(0))\leq \is(H(P))$ and $\is(H(0))\leq \is(H)$. See Corollary \ref{m3}. 

{\bf (6) Measures associated with ground states:} \\
In \cite{fr1,fr2} 
spectral properties of the translation invariant model 
including the Nelson model and the Pauli-Fierz model are investigated, 
in which mainly the renormalized Nelson model 
with nonrelativistic or relativistic kinematic term is studied.
See also  \cite {ca, fgs, lo1,lo2, gelo,ts}. 
In \cite{fr1},  it is shown that 
a ground state of the fiber Hamiltonian of the Pauli-Fierz model exists for all values of coupling constants but 
$|P|<P_0$ with some $P_0$ for a massive case. 
In \cite{ch}, it is extended to a massless case.
Although the existence problem of ground states mentioned above is solved, 
it is not constructive. 
In \cite{bhlms} functional integrals are applied to study properties of 
ground state $\gr$ of the Nelson model, in which $(\gr, {\cal O}\gr)_{{\hhh}_N}$ with 
suitable operator ${\cal O}$  
is represented as 
\eq{bhlms}
(\gr, {\cal O}\gr)_{\hhh_N}=\int_{C(\RR;\BR)} f_{\cal O}(q) d\mu_\infty(q)
\en 
with some function $f_{\cal O}$ and a  probability measure $d\mu_\infty$ 
on  $C(\RR;\BR)$. 
This measure is constructed by taking an infinite time limit 
of the form \kak{FFFFF}.  
In this paper, we do not construct such a measure, 
since it is not easy to control the stochastic integral appeared in \kak{st}.
Instead of this, as is studied in \cite[Theorem 3.4.1]{glja}, 
we construct a sequence 
of measures 
$$\{e^{iP\cdot b(2t)}d\mu_{2t}\}_{t>0}$$ converging, in some sense, to $(\gr(P), {\cal O}\gr(P))_\fffb$ 
with a ground state $\gr(P)$ of $H(P)$. 
Actually due to a double stochastic integral it has {\it informally}
 expressed as 
\eq{2t}
d \mu_{2t} =
\frac{1}{Z}\exp\lk 
-\frac{e^2}{4}
\sum_{\alpha,\beta=1}^ d
\int_0^{2t}db_\alpha(s)\int_0^{2t}db_\beta(s')
W_{\alpha\beta}(s-s',b(s)-b(s'))\rk db
\en 
and 
$$W_{\alpha\beta}(t,x)=
\int_\BR (\delta_{\alpha\beta}-\frac{k_\alpha k_\beta}{|k|^2}) 
\frac{|\vp(k)|}{\omega(k)}e^{-|s|\omega(k)}e^{-ikx}dk.$$
See Corollary \ref{ex} and Remark \ref{exx} for a detail. 
The properties of 
measure $d\mu_{2t}$ with $\int db_\alpha(s)\int db_\beta(s')$ replaced by $\int ds\int ds'$, 
which corresponds to the measure associated with the ground state of the Nelson model, 
is 
discussed in \cite{betz, belo, lomi}.
We shall discuss 
the existence of measures such as \kak{2t} on a continuous path space in \cite{tight}.

\subsection{Remarks and plan of the paper}
Recently the spectral properties of a general version of 
the Pauli-Fierz model with a fixed total momentum is studied  in \cite{lms} where 
the self-adjointness and energy inequalities  are  also shown. 
See also \cite{jsm1, jsm2} for some recent development for the massive Nelson model, 
and \cite{Ar,agg,agg2,ag, sasa} for a relativistic model.  
The effective mass $m_{\rm eff}$ is defined 
by the inverse of the Hessian of the ground state energy $E(P)$ of 
a fiber Hamiltonian $H(P)$ at $P=0$, i.e., 
$m_{\rm eff}^{-1}=\partial^2 E(P)/\partial|P|^2\lceil_{P=0}$. 
For the effective mass of the Pauli-Fierz model without infrared cutoff 
is studied in \cite{ch2, bcfs}, and its renormalization in e.g., 
\cite{hs, hk,  hisp3, lilo}. In this paper we do not discuss a relationship between 
the effective mass and functional integrals. 
See \cite{vo,sp5} to this direction for the Nelson model. 
See \cite{sp1} as a review of the recent development on this area. 

This paper is organized as follows. 
In Section 2 we define the Pauli-Fierz Hamiltonian $H(P)$ for an arbitrary 
 total momentum 
$P\in\BR$ and an arbitrary coupling constant, 
and discuss a relationship between rotation invariance  
and polarization vectors. 
Moreover we introduce an operator $K(P)$ defined by $H(P)$ 
without the free Hamiltonian $\hf$ of the field. 
In Section~3 we construct a functional integral representation of 
$(\Psi, \ehtp\Phi)_\fffb$ and 
show some applications including the diamagnetic inequality, 
the positivity improvingness of 
$\t e^{-tH(0)}\tf $ and the essential self-adjointness of $K(P)$. 
Section~4 is devoted to extending  the functional integral representation to 
an $n$ point Euclidean Green function and to giving  applications.

\section{The Pauli-Fierz Hamiltonian}
\subsection{Preliminaries and notations}
Let us assume that an electron moves in the $d $ dimensional space and is polarized to 
$d -1$ directions. 
Let $\fff$ be the Boson Fock space over
${\cal W}:=\oplus^{d-1}\LR$, i.e.,
$$\fff:=\bigoplus_{n=0}^\infty \fff^{(n)}:=
\bigoplus_{n=0}^\infty
[\otimes_s^n {\cal W}] ,$$
where $\otimes_s^n {\cal W}$ denotes the $n$-fold symmetric tensor product of
Hilbert space ${\cal W}$, i.e., $\otimes_s^n {\cal W} :=S_n(\otimes^n W )$
with $\otimes_s^0{\cal W}:=\CC$.
 Here
$S_n$ symmetrizes $\otimes^n W $, i.e.,
$$\d S_n(f_1\otimes\cdots\otimes f_n):=
\frac{1}{n!}\sum_{\s\in \wp_n} f_{\s(1)}\otimes \cdots \otimes 
f_{\s(n)},$$
where $\wp_n$ denotes the set of permutations of degree $n$.
 In this paper we denote the norm and the scalar product on a Hilbert space 
${\cal K}$ by
$\|f\|_{\cal K}$ and $(f,g)_{\cal K}$, respectively.
The scalar product is linear in $g$ and antiliner in $f$.
Unless confusions arise we omit ${\cal K}$. 
$\fff$ can be identified with the set of $\ell_2$-sequences 
$\{\Psi^{(n)}\}_{n=0}^\infty $ with $\Psi^{(n)}\in\fffb^{(n)}$ 
such 
that
$\sum_{n=0}^\infty\|\Psi^{(n)}\|_{\fff^{(n)}}^2<\infty$ and $\fff$ is the 
Hilbert space endowed with the scalar product
$(\Psi, \Phi)_\fffb=\sum_{n=0}^\infty (\Psi^{(n)},\Phi^{(n)})_{\fff^{(n)}}$.
$\Omega=\{1,0,0,...\}\in\fff$ is called as the Fock vacuum.
The annihilation operator and the creation operator on $\fff$
are denoted by $a(f)$ and $\add(f)$, $f\in W$, respectively, which are 
defined by
$$(\add(f)\Psi)^{(n)}:=\sqrt n S_n(f\otimes \Psi^{(n-1)})$$
with the domain
$$D(\add(f)):=\{\{\Psi^{(n)}\}_{n=0}^\infty \in\fff| \sum_{n=1}^\infty n\|
S_n(f\otimes \Psi^{(n-1)})\|_{\fff^{(n)}}^2<\infty\},
$$
and $a(f)=(\add(\bar f))^\ast$. 
Since the creation operator and the annihilation operator are closable, 
we take their closed extension and 
denote them by the same symbols. 
Let
$\ffff$ be the so-called finite particle subspace of $\fffb$ defined by
$$\ffff:=\{\{\Psi^{(n)}\}_{n=0}^\infty \in\fff| \Psi^{(m)}=0 \mbox{ for all } m \geq 
\exists M\}.$$
The annihilation operator and the creation operator leave $\ffff$ invariant 
and satisfy the canonical 
commutation relations on it:
$$[a(f),\add(g)]=(\bar f,g)1,\ \ \ [a(f),a(g)]=0,\ \ \ [\add(f), 
\add(g)]=0.$$
For $f=(f_1,...,f_{d -1})\in\oplus^{d-1}\LR$,   we informally write $\ass(f)$, where 
$\ass$ stands $a$ or $\add$,  as
$ \d \ass(f)=\jjj \int \ass(k,j) f_j(k) dk$ 
with informal kernel $\ass(k,j)$.  
Let $T$ be a contraction operator  on $\LR$.
Then the  contraction linear operator $\Gamma (\K T_ {d-1})$ on $\fff$ is defined by
$$\Gamma(\K T_{d-1}):=\bigoplus_{n=0}^\infty \otimes ^n \K T_ {d-1},$$
where
$\K T_\ell  := \underbrace{T\oplus\cdots\oplus T}_{\ell}$. 
Unless confusions  arise we  write $\Gamma(T)$ for 
$\Gamma(\K T_ {d-1})$.
For a self-adjoint operator $h$ on ${\cal W}$, 
$\{\Gamma (e^{it h})\}_{t\in\RR}$ is a strongly continuous one-parameter 
unitary group on $\fff$.
Then by the Stone theorem \cite{rs2}, 
there exists a unique self-adjoint operator $\dg h$ on 
$\fff$ such that 
$$\Gamma(e^{it h})=e^{it\dg h},\ \ \ t\in\RR.$$
$\dg h$ is called as the second quantization of $h$. 
For a self-adjoint operator $h$ in $\LR$, 
$\dg{[h]_{d-1}}$ is simply denoted by $\dg h$ unless confusion arises.  
The number operator is defined by $N:=\dg 1$.
Let 
$$\omega(k)=|k|$$ be the multiplication operator on $\LR$. 
Define the free Hamiltonian $\hf$ on $\fff$ by
$\hf:=\dg \omega$. 
The quantized radiation field $\av_\mu(x)$, $x\in \BR$, $\mu=1,...,d $,
with a form factor $\varphi$  is defined by
\begin{eqnarray*}
\av_\mu(x)
=\frac{1}{\sqrt 2}  \jjj\int e_\mu 
(k,j)(
\frac{\vp(k)}{\sqrt{\omega(k)}} \add(k,j) e^{-ik\cdot x}+
\frac{\vp(-k)}{\sqrt{\omega(k)}}   a(k,j) e^{ik\cdot x}) dk, 
\end{eqnarray*}
which acts on $\fffb$. 
Here $e(k,1),\cdots, e(k,d -1)$ denote  generalized polarization vectors satisfying
$k\cdot e(k,j)=0$ and $e(k,i)\cdot e(k,j)=\delta_{ij} 1$, $i,j=1,...,d-1$, 
and  
$\vp$  is the Fourier transform of form factor $\varphi$ given by
$\d \vp(k)=(2\pi)^{-d/2} \int _\BR \varphi(x) e^{-ik\cdot x} dx$.
Note that 
$$\sum_{j=1}^{d -1} e_\alpha (k,j) e_\beta (k,j)=\delta_{\alpha\beta }-\frac{k_\alpha  k_\beta }{|k|^2}
:=
\dep_{\alpha\beta}(k),\ \ \
\alpha, \beta=1,...,d. $$
Throughout this paper we assume (A) below. 

{\bf (A)}
{\it Form factor $\vp$ satisfies that $\sqrt\omega\vp, \vp/\omega\in\LR$ and
$\ov {\vp(k)}=\vp(-k)=\vp(k).$}

$\av_\mu(x)$ is essentially self-adjoint on $\ffff$, and   
its unique self-adjoint extension is denoted by the same symbol. 
The Hilbert space $\hhh$ of state vectors for the total  system under consideration is given 
by the tensor product Hilbert space: 
$$\hhh:=L^2(\RR^d _x)\otimes\fff.$$
Under the identification of $\hhh$ with the set of $\fff$-valued 
$L^2$-functions on $\BR$, i.e.,
\eq{esi11}
\hhh\cong \int_\BR^\oplus \fff dx,
\en  we define the self-adjoint operator $\av$ on ${\cal H}$ by  
$$\av_\mu :=\int_\BR^\oplus \av_\mu (x) dx,\ \ \ \mu=1,...,d ,$$
i.e.,
$(\av_\mu F)(x) = \av_\mu (x) F(x)$ 
and 
$$D(\av_\mu):=\{F\in\hhh|F(x)\in D(\av_\mu(x))\mbox{ a.e. } x\in \BR \mbox { and } 
\int_\BR\|\av_\mu(x) F(x)\|_\fffb^2 dx<\infty\}.$$
We set 
$\av=(\av_1,...,\av_d )$.
From the fact that $k\cdot e(k,j)=0$ we have  
$$\nabla\cdot\av=\sum_{\mu=1}^d [\nabla_\mu, \av_\mu]=0$$ as an operator. 
We use the identification \kak{esi11} without notices.
We define the decoupled self-adjoint Hamiltonian $H_0$ by
$$
H_0:=(-\frac{1}{2}\Delta+V)\otimes 1+1\otimes \hf,\quad 
D(H_0):=D(-\Delta\otimes 1)\cap D(1\otimes\hf).
$$
The total Hamiltonian $H$, the so-called Pauli-Fierz Hamiltonian,  is described by the minimal coupling, 
$-i\nabla_\mu\otimes 1\rightarrow -i\nabla_\mu\otimes 1-e\av_\mu$, 
to $H_0$. Then 
$$H:=\frac{1}{2} (-i\nabla\otimes 1-e\av)^2+V\otimes 1+1\otimes\hf,$$
where $e\in\RR$ is a coupling constant. 
As a mathematical interest we introduce another operator $K$ by  
$$K:=\frac{1}{2}(-i\nabla\otimes 1-e \av)^2+V\otimes 1.$$

It is well known that 
\eq{aaa}
\|\ass(f)\phi\|\leq 
\|f/\sqrt\omega\| \|\hf^\han\Psi\|+\|f\|\|\Psi\|,\quad \Psi\in D(\hf^\han).
\en 
Assumption $\ov {\vp(k)}=\vp(-k)=\vp(k)$ implies that $H$ is symmetric, 
and $\sqrt\omega\vp, \vp/\omega\in\LR$ 
that $(-i\nabla\otimes1)\av+\av(-i\nabla\otimes 1)$ and $\av^2$ are 
relatively bounded with respect to $-\Delta\otimes 1+1\otimes \hf$. 
The proposition below is established in \cite{h11,h16}. 
\bp{hi123}
Assume  that
$V$ is relatively bounded with respect to $-\Delta$ with a relative bound
strictly smaller than one. 
Then 

(1) $H$ is self-adjoint  on $D(H_0)$ 
and essentially self-adjoint on any core of self-adjoint operator
$-(\han) \Delta\otimes 1 +1\otimes \hf$, and bounded from below, 

(2)  $K$ is essentially self-adjoint on $C^\infty(\Delta\otimes 1)\cap 
C^\infty(1\otimes N)$ and bounded from below, where $C^\infty(T):=\cap_{n=1}^\infty D(T^n)$.
\ep 
We denote the self-adjoint  extension of 
$K\lceil_{C^\infty(\Delta\otimes 1)\cap 
C^\infty(1\otimes N)}$ by the same symbol $K$. 
Throughout this paper we assume that $V$satisfies the same assumptions 
in Proposition~\ref{hi123}.

\subsection{Translation invariance}
In this subsection we set $V=0$.
Define the field momentum by
$\pf_\mu :=\dg {k_\mu}$, $\mu=1,...,d$, 
and
the total momentum  
$${\tot}_\mu:=\ov{-i\nabla_\mu \otimes 1+1\otimes\pf_\mu},\ \  \ \mu=1,...,d ,$$
and set
$\pf:=(\pf_1,...,\pf_d )$, 
$\tot:=({\tot}_1,...., {\tot}_d )$, where $\ov{X}$ denotes the closure of closable operator $X$. 
It is seen that $H$ is translation invariant \cite[(5.23)]{h11}, i.e., 
$$e^{is {\tot}_\mu} He^{-is {\tot}_\mu}=H,\ \ \ s\in\RR,\ \ \ \mu=1,...,d .$$
We shall decompose $H$ on $\s(\tot_\mu)=\RR$.
Operator $H(P)$, $P\in\BR$,  acting in $\fffb$ is defined by 
$$
H(P):=\frac{1}{2}(P-\pf-e\av(0))^2+\hf,\quad 
D(H(P)):= D(\hf)\cap D(\pf^2).
$$
Note that $H(P)$ is a well defined symmetric operator 
on $D(\hf)\cap D(\pf^2)$ by assumption {\bf (A)}. 
For a sufficiently small $e$, 
the self-adjointness of  $H(P)$ is easily shown by using 
\kak{aaa} and the Kato-Rellich theorem. 
In order to show the self-adjointness of $H(P)$ for an arbitrary $e\in\RR$, 
we need to make a detour.

\bt{po}
$H(P)$ is self-adjoint and 
\eq{int1}
 \int^\oplus_{\BR}H(P) dP\cong H.
\en 
\et
Although it is not a physically reasonable model, 
it is of interest to study 
the essential self-adjointness of another operator $K(P)$ defined by 
$$
K(P):=\frac{1}{2}(P-\pf-e\av(0))^2,\ \ \ 
D(K(P)):=D(\pf^2)\cap D(\hf).
$$
It is not clear that $K(P)$ is self-adjoint even for a sufficiently small $e$ by the lack of $\hf$. 
The quadratic form $\tilde Q_P(\Psi,\Phi)$ is given by 
\begin{eqnarray*}
&&
\tilde Q_P(\Psi, \Phi):=\frac{1}{2}\sum_{\mu=1}^d 
((P-\pf-e\av(0))_\mu \Psi, (P-\pf-e\av(0))_\mu \Phi)_\fffb, \\ 
&&D(\tilde Q_P):=\cap_{\mu=1}^ d [D(\pf_\mu)\cap  D(\av(0)_\mu)].
\end{eqnarray*}
Since $\tilde Q_P$ is a densely defined nonnegative quadratic form, 
there exists a  positive self-adjoint operator $K_{\rm F}(P)$ such that 
$\tilde Q_P(\Psi, \Phi)=(K_{\rm F}(P)^\han\Psi, K_{\rm F}(P)^\han\Phi)$.

\bt{poo}
(1) It follows that  
\eq{int2}
 \int^\oplus_{\BR}K_{\rm F}(P) dP\cong K.
\en 
(2)  Assume that $\omega^{3/2}\vp\in\LR$. 
Then $K(P)$ is 
essentially self-adjoint and 
\eq{int3}
 \int^\oplus_{\BR}\ov{K(P) }dP\cong K.
\en 
\et
Theorem \ref{poo} (2) is proved by 
using a functional integral representation in Section~\ref{332}. 
We here prove Theorem \ref{po} and Theorem \ref{poo} (1). 

The fiber decomposition of $H$ will be achieved 
through the unitary operator \\
$U:L^2(\RR^d _x)\otimes \fff
\rightarrow L^2(\RR_\xi^d )\otimes\fff$ 
given by
\eq{esi12}
U:=({\Bbb F}\otimes 1) 
\int_{\BR}^\oplus
\exp\lk
i x\cdot \pf \rk  dx,  
\en
where ${\Bbb F}:L^2(\RR^d _x)\rightarrow L^2(\RR^d _\xi)$ denotes 
the Fourier transformation on $\LR$. 
Actually for 
$\Psi\in L^2(\RR^d _x)\otimes \fff$,
$$(U\Psi)(\xi)=
\frac{1}{\sqrt{(2\pi)^d }}
\int_\BR e^{-ix\cdot\xi}e^{i  x\cdot \pf} 
\Psi(x) dx,$$
where $\int\cdots dx$ denotes  the $\fff$-valued integral in the strong topology.
Let $Q_\mu$ denote the multiplication operator in $\LR$ defined by 
$(Q_\mu f)(\xi):=\xi_\mu f(\xi)$, $\mu=1,...,d$. 
Set  
$$\fffb_\infty:=\mbox{ L.H. } \{\add(f_1)\cdots 
\add(f_n)\Omega|
f_j\in   C_0^\infty(\BR), j=1,...,n, n=1,2,..\}\cup\{{\Bbb C}\Omega\}$$
 and 
$\cd:=C_0^\infty(\BR)\hat\otimes \fffb_\infty$, 
where $\mbox{L.H.}\{\cdots\}$ denotes the linear hull of $\{\cdots\}$ and 
$\hat\otimes$ is the algebraic tensor product, i.e., 
the set of vectors of the form 
$\sum_{j=1}^{\rm finite}\alpha_j f_j\otimes \phi_j$, $\alpha_j\in{\Bbb C}$, $f_j\in C_0^\infty(\BR)$ and 
$\phi_j\in\fff_\infty$.
We define $L$ by 
$$
L:=\ov{
\left. 
\frac{1}{2}(Q\otimes 1-1\otimes \pf-e 1\otimes 
\av(0))^2+1\otimes \hf \right\lceil_{\cd}}. 
$$

\bl{Gr}(1) $L$ is self-adjoint on $D((\ov{\QPP})^2)\cap D(1\otimes \hf))$,  \\
(2) $U H U\f =L$ 
on $D((\ov{\QPP})^2)\cap D(1\otimes \hf))$. 
\el

\proof 
Not that  
$e^{ix\cdot \pf}\add(f) e^{-ix\cdot \pf}= \add(e^{ik\cdot x} f)$ and 
$e^{ix\cdot \pf}a (f) e^{-ix\cdot \pf}= a (e^{-ik\cdot x} f)$. 
Hence we have 
$UH\Phi=LU\Phi$ for $\Phi\in\cd$. 
Since $\cd$ is a core of $H$ and  $L$ is closed, 
we obtain that $U$ maps $D(H)$ onto $D(L)$ with $UHU\f=L$. Then $L$ is self-adjoint on $UD(H)$. 
Since $U D(-\Delta\otimes 1)=D((\ov{\QPP})^2)$ and $UD(1\otimes\hf)=D(1\otimes\hf)$, 
we have 
$
UD(H)=
D ((\ov{\QPP})^2)\cap D(1\otimes \hf)$.   
Thus (1) and (2) follow. 
\qed

{\it {\bf Proof of Theorem \ref{po} and Theorem \ref{poo} (1)}}

\proof 
The quadratic form $Q_P(\Psi,\Phi)$ is given by 
\begin{eqnarray*}
&&
\hspace{-1cm}
Q_P(\Psi, \Phi):=\frac{1}{2}\sum_{\mu=1}^d 
((P-\pf-e\av(0))_\mu \Psi, (P-\pf-e\av(0))_\mu \Phi) 
+(\hf^\han\Psi, \hf^\han\Phi),\\
&&
\hspace{-1cm}
D(Q_P):=\cap_{\mu=1}^ d [D(\pf_\mu)\cap D(\av(0)_\mu)]\cap D(\hf^\han).
\end{eqnarray*}
Since $Q_P$ is a densely defined nonnegative quadratic form, 
there exists a positive self-adjoint operator $\ffri(P)$ such that 
$Q_P(\Psi, \Phi)=(\ffri(P)^\han\Psi, \ffri(P)^\han\Phi)$. 
Define the self-adjoint operator $\tilde H$ acting in $\hhh$ by  
$\d \tilde H:=\int^\oplus _{\BR} \ffri(P) dP$. 
For $\phi\in \cd$, 
$\d U\phi(P)={(2\pi)^{-d/2}}
\int_{\BR} e^{-ix\cdot P}e^{ix\cdot\pf}\phi(x) dx\in D(\hf)\cap D(\pf^2).
$
Then for $\psi,\phi\in\cd$, we have 
\begin{eqnarray*}
 (U\psi, LU\phi)_{\hhh}
&=& \int_{\BR} dP (U\psi(P), [\frac{1}{2}(P-\pf-e\av(0))^2+\hf] U\phi(P))_\fffb \\
&=& \int_{\BR}  dP (U\psi(P), \ffri(P) U\phi(P))_\fffb 
=(U\psi, \tilde H U\phi)_\hhh.
\end{eqnarray*}
Hence 
$U\f LU=U\f \tilde H U$ on $\cd$ and then $H=U\f \tilde H U$ on $\cd$ by Lemma \ref{Gr}.  
Since $\cd$ is a core of $H$ and $\tilde H$ is  self-adjoint, 
we can see that $U$ maps $D(H)$ onto $D(\tilde H )$ with $UHU\f=\tilde H$. Then 
$\d \int_\BR^\oplus \ffri(P) dP\cong H$ is obtained. 
The proof of Theorem~\ref{poo} (1) 
is similar.  
The proof of self-adjointness of $\ffri(P)$ below is 
due to \cite{lms}.
In \cite{h11,h16} it is proved that 
there exists a constant $C$ such that 
\eq{k121}
\|H_0F\|_\hhh\leq C\|(H+1)F\|_\hhh,\ \ \ F\in D(H).
\en 
We define $\fri_0(P)$ by $\fri_0(P):=\frac{1}{2}(P-\pf)^2+\hf$ and $D(H_0(P))=D(\hf)\cap D(\pf^2)$.  
Note that $\fri_0(P)$ is self-adjoint and 
$\d \int_\BR^\oplus H_0(P) dP=H_0$. 
Then we have for $F\in\hhh$ such that $(UF)(P)=f(P)\Phi$ with 
$f\in C_0^\infty(\BR)$ and $\Phi\in\fffb_\infty$ by \kak{k121}, 
$$
\int_\BR |f(P)|^2 \|\fri_0(P)\Phi\|_\fffb^2 dP
\leq C^2\int_\BR f(P)^2 \|(\ffri(P)+1)\Phi\|_\fffb^2 dP.
$$
Here since $f\in C_0^\infty(\BR)$ is arbitrary, we see that 
\eq{K2}
\|\fri_0(P)\Phi\|_\fffb \leq 
C \|(\ffri(P)+1)\Phi\|_\fffb
\en 
for almost everywhere $P\in\BR$. 
Since the both-hand sides of \kak{K2} are  continuous in $P$, 
\kak{K2} holds for all $P\in\BR$. 
\kak{K2}  implies that $\fri_0(P)(\ffri(P)+1)^{-1}$ is bounded and then  
$H_0(P)e^{-t\ffri(P)}$ is bounded, which implies that $e^{-t\ffri(P)}$ 
leaves $D(H_0(P))=D(\hf)\cap D(\pf^2)$ invariant. Then 
$\ffri(P)$ is essentially self-adjoint on $D(\hf)\cap D(\pf^2)$ by \cite[Theorem X.49]{rs2}. 
Moreover \kak{K2} yields that  $\ffri(P)$ is closed on  $D(\hf)\cap D(\pf^2)$. 
Hence $\ffri(P)$ is self-adjoint on $D(\hf)\cap D(\pf^2)$. 
By the fundamental inequality derived by \kak{aaa}:
$$\|\fri_{\rm F}(P)\Phi\|_\fffb\leq C_1\|H_0(P)\Phi\|_\fffb+C_2\|\Phi\|_\fffb,\ \ \ \Phi\in D(H_0),$$
we can see that 
$\ffri(P)$ is essentially self-adjoint on any core of $H_0(P)$, where $C_1$ and $C_2$ are some constants. 
Since $H(P)=\ffri(P)$ on $D(\hf)\cap D(\pf^2)$, Theorem \ref{po} follows. 
Then the proof is complete. 
\qed

\bc{c3}
Let $\Lambda>0$  and 
$\d \vp_\Lambda (k):=\lkk\begin{array}{ll} 1/\sqrt{(2\pi)^3},& |k|<\Lambda,\\
0,&|k|\geq \Lambda.\end{array}\right.$ 
Then $H(P)$ with $\vp$ replaced by $\vp_\Lambda$ is self-adjoint for arbitrary $P\in\BR$, $e\in\RR$ and $\Lambda>0$. 
\ec
\proof 
Since $\vp$ satisfies (A), the corollary follows. 
\qed

\subsection{Rotation invariance, helicity and degeneracy of ground states}
In this subsection we discuss the rotation invariance of $H$ and $H(P)$. 
For simplicity we set $d=3$, and add a spin to $H$ and $H(P)$. Namely 
let 
$\s=(\s_1,\s_2,\s_3)$ be  $2\times 2$ Pauli matrices such that 
$\s_\mu\s_\nu+\s_\nu\s_\mu=2\delta_{\mu\nu}$.
We define two operators $\hs$ and $\hs(P)$ acting on ${\Bbb C}^2\otimes \hhh$ and ${\Bbb C}^2\otimes\fffb$, 
respectively, by 
\begin{eqnarray*}
&& \hs:=1\otimes H+\sum_{\mu=1}^3 \s_\mu\otimes  \Hs_\mu,\\
&& \hs(P):=1\otimes H(P)+\sum_{\mu=1}^3 \sigma_\mu\otimes  \Hs(0)_\mu,
\end{eqnarray*}
where 
$\d \Hs_\mu:=-\frac{e}{2}\int_{\RR^3}^\oplus B_\mu(x) dx$ 
with $B(x):={\rm rot}_x\av(x)$,  
and 
$\d \Hs_\mu(0):=-\frac{e}{2} B_\mu(0).$
In the similar way to the proof of Theorem \ref{po} 
it can be shown that $\hs$ and $\hs(P)$ 
are self-adjoints operator on ${\Bbb C}^2\otimes D(H)$ 
and ${\Bbb C}^2\otimes [D(\pf^2)\cap D(\hf)]$, respectively. 
Let $R\in SO(3)$ and $\hat k=k/|k|$. 
The relationship between two orthogonal bases $e(Rk,1),e(Rk,2), \hat {Rk}$ and 
$Re(k,1),Re(k,2), R\hat {k}$ in $\RR^3$ at $k$ is as follows:
\eq{pol3}
\vvv {e(Rk,1)\\ e(Rk,2)\\ \hat{Rk}} =
\MMMM{\cos\theta 1_3}{-\sin\theta 1_3}{0}{\sin\theta 1_3}{\cos\theta 1_3}{0} 0 0 {1_3}
\vvv{Re(k,1)\\Re(k,2)\\ R\hat k},
\en 
where $1_3$ denotes the $3\times 3$ identity matrix and 
\eq{ka}
\theta:=\theta(R,k):=\arccos(Re(k,1)\cdot e(Rk,1)).
\en 
Let  $R=R(\phi,n)\in SO(3)$  
be  the rotation around $n\in S^2:=\{k\in\RR^3||k|=1\}$ 
with angle $\phi\in\RR$ and ${\rm det} R=1$.  
Let $\llk:=k\times (-i\nabla_k)=(\llk_1,\llk_2,\llk_3)$ 
be the triplet  of angular momentum operators in $L^2(\RR_k^3)$. 
Then 
\eq{f1}
e^{i\theta(R,k) X}e^{i\phi n\cdot \llk}
\vvv {e(k,1)\\ e(k,2)}=\vvv {R e(k,1)\\ R e(k,2)},
\en 
where 
$$X=-i \MMM  0 {-1_3} {1_3}  0 
:\begin{array}{ccc}\RR^3\oplus\RR^3&\longrightarrow  &\RR^3\oplus\RR^3\\
x\oplus y &\longmapsto  &
-i(-y\oplus x).\end{array}$$
In general, the polarization vectors of photons with momentum $k$  is arbitrary given, but form a right-handed system at $k$. 
To discuss a rotation symmetry of $\hs$ and $\hs(P)$, 
we introduce {\it coherent} polarization  vectors to some direction.
Assumption (P) is as follows. 

{\bf (P)} {\it There exists $(n, w)\in S^2\times{\Bbb Z}$ 
such that 
polarization vectors $e(\cdot, 1)$ and $e(\cdot,2)$ satisfy for 
$R=R(n,\phi)\in SO(3)$  and for $\hat k\not =n$, 
\eq{pol}
\vvv {e(Rk,1)\\ e(Rk,2)}=\MMM{\cos(\phi w)1_3}
{-\sin( \phi w)1_3}{\sin( \phi w)1_3}
{\cos(\phi w)1_3}\vvv {R e(k,1)\\ R e(k,2)},\ \  \phi\in\RR.
\en
 }
Assume (P). Then 
\eq{pol2}
e^{i\phi (w X+n\cdot \llk)}
\vvv { e(k,1)\\  e(k,2)}
=\vvv {R e(k,1)\\ R e(k,2)}.
\en
We show some examples for polarization vectors satisfying (P). 

\begin{example}\label{e1}
Let $n_z=(0,0,1)$. Given polarization vectors 
$e(\hat k_0,1)$ and $e(\hat k_0,2)$ 
for $\hat k_0\in {\Bbb S}=\{(\sqrt{1-z^2}, 0, z)\in S^2|-1\leq z\leq 1\}$. 
For $\hat k=(\hat k_1,\hat k_2,\hat k_3)$,  
there exists $0\leq \phi<2\pi$ such that $R(n_z,\phi)\hat k_0=\hat k$, where 
$k_0=(\sqrt{1-{\hat k}_3^2}, 0, \hat k_3)\in{\Bbb S}$.  
Define 
\eq{polp2}
\vvv {e(k,1)\\ e(k,2)}:=\MMM{\cos \phi1_3}
{-\sin\phi1_3}{\sin \phi1_3}
{\cos \phi 1_3}\vvv {R(n_z,\phi) e(\hat k_0,1)\\ 
R(n_z,\phi) e(\hat k_0,2)}.
\en
It is  checked that $e(k,j)$ satisfies 
\kak{pol} with $(n_z, 1)\in S^2\times{\Bbb Z}$.
\end{example}

\begin{example}\label{lilo}
Let $n\in S^3$ and $e(k,1):=\hat k\cdot n/\sin\theta$ and $e(k,2):=(k/|k|)\times e(k,2)$, 
where $\theta=\arccos(\hat k\cdot n)$. 
Then $e(k,j)$ satisfies \kak{pol} with $(n,0)\in S^2\times{\Bbb Z}$.
\end{example}
Assume (P) with some $(n,w)\in  S^2\times{\Bbb Z}$. 
We define 
$S_{\rm f}:=d\Gamma(w {\Bbb X})$ and 
$L_{\rm f}:=d\Gamma(\llk)$.
Here 
\eq{xxx}
{\Bbb X}:=-i \MMM  0 {-1} {1}  0 :
\begin{array}{ccc}
L^2(\RR^3)\oplus L^2(\RR^3)&\longrightarrow & L^2(\RR^3)\oplus L^2(\RR^3)\\
f\oplus g& \longmapsto & -i(-g\oplus f).\end{array}
\en 
$S_{\rm f}$ is called the 
helicity of the field\footnote{It is  written informally 
as  
$\d S_{\rm f}=i\int w  (\add(k,2)a(k,1)-\add(k,1)a(k,2)) dk$} 
and $L_{\rm f}$ the angular momentum of the field. 
Define $J_{\rm f}$ and $J_{\rm p}$ by 
$$
J_{\rm f}:=n\cdot L_{\rm f}+S_{\rm f},\quad 
J_{\rm p}:=n\cdot \llx+\half n\cdot \s 
$$
and set 
$$J_{\rm total}:=J_{\rm p}\otimes 1 + 
1 \otimes J_{\rm f}$$
which 
acts 
in $L^2(\RR^3;{\Bbb C}^2)\otimes \fff$.
\bl{rot1}
Assume (P) and that  
$\vp(R(n,\phi)k)=\vp(k)$ for $\phi\in\RR$.   
Then $\hs$ is rotation invariant around $n$, i.e., 
\eq{pol123}
e^{i\phi J_{\rm total}} \hs 
e^{-i\phi J_{\rm total}}=\hs,\ \ \ 
\phi\in\RR.
\en 
\el
\proof 
By $e^{i\phi J}=e^{i\phi  S_{\rm f}}
e^{i\phi n\cdot L_{\rm f}}$ and \kak{pol2}, 
we see that 
\begin{eqnarray}
&&\label{1234}
 e^{i\phi J_{\rm f}}  \hf 
e^{-i\phi J_{\rm f}}=\hf,\\
&&
\label{a12345}
 e^{i\phi  J_{\rm f}} \pf_\mu 
e^{-i\phi J_{\rm f}} =(R(n,\phi)\pf)_\mu,\\
&&
\label{123456}
  e^{i\phi  J_{\rm f}} \av_\mu(x)
e^{-i\phi J_{\rm f}}=
(R(n,\phi)\av(R(n,\phi)^{-1}x))_\mu,\\
&&
\label{1234567}
  e^{i\phi J_{\rm f}} B_\mu(x)
e^{-i\phi J_{\rm f}}=
(R(n,\phi)B(R(n,\phi)^{-1}x))_\mu.
\end{eqnarray}
Since 
\begin{eqnarray*}
&& 
e^{i\phi n\cdot \llx} x_\mu
 e^{-i\phi n\cdot \llx}  =(R(n,\phi)x)_\mu, \\ 
&& 
e^{i\phi n\cdot \llx} (-i\nabla_x)_\mu 
 e^{-i\phi n\cdot \llx}  =(R(n,\phi)(-i\nabla_x))_\mu,\\
&& 
e^{i\phi n\cdot (\han)\s} \s_\mu  
 e^{-i\phi n\cdot (\han)\s} =(R(n,\phi)\s)_\mu,
\end{eqnarray*}
we can see \kak{pol123} by \kak{1234}-\kak{1234567}.
\qed
Note that $\s(n\cdot(\llx+(\han)\s))={\Bbb Z}_\han$, 
$\s(n\cdot L_{\rm f})={\Bbb Z}$ and 
$\s(S_{\rm f})={\Bbb Z}$, 
since 
\eq{s123}
\s(\dg h)=\ov{
\{0\}\bigcup 
\cup_{n=1}^\infty \{\lambda_1+\cdots+\lambda_n|\lambda_i\in\s(h),j=1,..,n\}}.
\en
Then $\s(J_{\rm total})={\Bbb Z}_\han$ and we have the theorem below.
\bt{rotdis}
We assume the same assumptions as in Lemma  \ref{rot1}.
Then  
$${\Bbb C}^2\otimes \hhh=\bigoplus_{z\in{\Bbb Z}_\han}\hhh(z),\ \ \ \ 
\hs=\bigoplus_{z\in {\Bbb Z}_\han}\hs(z).$$
Here $\hhh(z)$ is the subspace of ${\Bbb C}^2\otimes\hhh$ spanned by eigenvectors of 
$J_{\rm total}$ with 
eigenvalue $z\in\z$ and $\hs(z)=\hs\lceil_{\hhh(z)}$.
\et
\proof This follows from Lemma \ref{rot1} and the fact that $\s(J_{\rm total})=\z$. 
\qed

Next we study $\hs(P)$. 
\bl{rot2} Let $\vp(k)$ be rotation invariant. 
Then for arbitrary polarization vectors, 
$\hs(P)$  is unitarily equivalent to $\hs( R^{-1} P)$ for 
arbitrary $R\in SO(3)$.
\el
\proof 
It is enough to show the lemma for an arbitrary $R=R(m,\phi)$, $m\in S^2$ 
and $\phi\in\RR$. 
For arbitrary polarization vectors $e(\cdot,1)$ and  $e(\cdot,2)$, 
we define 
$h_{\rm f}=d\Gamma(\theta(R,\cdot){\Bbb X})$, 
where $\theta(R, k)=\arccos(e(Rk,1),Re(k,1))$ is given in \kak{ka}
and ${\Bbb X}$ in \kak{xxx}.
Thus we can see that 
\begin{eqnarray*}
&&\label{2234}
 e^{ih_{\rm f}}  e^{i\phi m\cdot L_{\rm f}}  \hf 
e^{-i\phi m\cdot L_{\rm f}}e^{-ih_{\rm f}} =\hf,\\
&&
\label{22345}
 e^{ih_{\rm f}}  e^{i\phi m\cdot L_{\rm f}}   \pf_\mu 
e^{-i\phi m\cdot L_{\rm f}}e^{-ih_{\rm f}} =(R(m,\phi)\pf)_\mu,\\
&&
\label{223456}
  e^{ih_{\rm f}}  e^{i\phi m\cdot L_{\rm f}}   \av_\mu(0)
e^{-i\phi m\cdot L_{\rm f}}e^{-ih_{\rm f}} =
(R(m,\phi)\av(0))_\mu,\\
&&
\label{2234567}
  e^{ih_{\rm f}}  e^{i\phi m\cdot L_{\rm f}}   B_\mu(0)
e^{-i\phi m\cdot L_{\rm f}}e^{-ih_{\rm f}} =
(R(m,\phi) B(0))_\mu.
\end{eqnarray*}
From these identities it follows that 
\eq{id}
e^{ih_{\rm f}}  e^{i\phi m\cdot ((\han)\s\otimes 1+1\otimes L_{\rm f})}  
\hs(P) 
e^{-i\phi m\cdot ((\han)\s)\otimes 1+1\otimes L_{\rm f})}e^{-ih_{\rm f}}
=\hs(R(m,\phi)^{-1}P).
\en 
Thus the lemma follows. 
\qed
Let $E_\s(P,e^2):=\is(H_\s(P))$. An immediate consequence of Lemma \ref{rot2} is as follows. 
\bc{ror} Let $\vp$ be rotation invariant. Then  $E_\s(RP,e^2)=E_\s(P,e^2)$ for arbitrary $R\in SO(3)$. 
\ec
\bt{rot4}
Assume (P) and 
that  
$\vp(k)=\vp(R(n,\phi)k)$ for $\phi\in\RR$. 
Then $\hs(P)$ is unitarily equivalent to $\hs(|P|n)$ and,   
${\Bbb C}^2\otimes \fffb$ and $\hs(|P|n)$ are  decomposed as  
\eq{sas}
{\Bbb C}^2\otimes \fffb=\bigoplus_{z\in \z} \fffb(z),\ \ \ \ 
\hs(P)\cong \hs(|P|n)=\bigoplus_{z\in \z} \hs(P,z).
\en
Here 
$\fffb(z)$ is the subspace spanned by eigenvectors of 
$n\cdot ((\han) \s\otimes 1+1\otimes J_{\rm f})$ with 
eigenvalue $z\in\z$ and 
$\hs(P,z)=\hs(P)\lceil_{\fffb(z)}$.
\et
\proof 
The fact $\hs(P)\cong \hs(|P|n)$ follows from Lemma \ref{rot2}. 
Since  
\eq{pn}
e^{i\phi  n\cdot ((\han)\s\otimes 1+1\otimes J_{\rm f})}\hs (|P|n)
e^{-i\phi  n\cdot ((\han) \s\otimes 1 +1\otimes J_{\rm f})}=
\hs(|P|n),\ \  \phi\in\RR, 
\en 
follows from \kak{id},  \kak{sas} is obtained. 
\qed

Let $\vp$ be rotation invariant and 
$\tilde\hs(P)$ be the Hamiltonian with polarization vectors 
given in Example \ref{lilo} with $n=n_z$, i.e., 
$$e(k,1)=(-k_2, k_1,0)/\sqrt{k_1^2+k_2^2},\quad 
e(k,2)=(k/|k|)\times e(k,1). 
$$
Then  Sasaki \cite{sasa} proved that 
$\hs(P)$ with arbitrary polarization vectors is 
unitarily equivalent to $\tilde\hs(P)$.
In particular  
$\hs(P)\cong \tilde\hs(P)=\bigoplus_{z\in {\Bbb Z}_{\han}}\tilde \hs(P,z)$. 
Moreover $\tilde\hs(P,z)\cong \tilde\hs(P,-z)$ for $z\in {\Bbb Z}_\han$. 
Let $M$ denote the multiplicity of ground state of $\hs(P)$. In \cite{hisp2},  the lower bound,  $M\geq 2$, is proven for a sufficiently 
small coupling constant. 
Sasaki \cite{sasa} gives an immediate consequence of $\tilde\hs(P,z)\cong \tilde\hs(P,-z)$. 
\bc{sasa2}
Let $\vp$ be rotation invariant. 
Then  $M$ is an even number. 
In particular $M\geq 2$, whenever ground states exist. 
\ec

Let us remove a spin  from $\hs(P)$.  
Assume (P) and that $\vp(R(n,\phi)k)=\vp(k)$ for $\phi\in\RR$.
Then, as is seen in \kak{pn}, 
as well as $\hs(|P|n)$,  $H(|P|n)$ is also rotation invariant around $n$, i.e., 
$$e^{i\phi n\cdot J_{\rm f}} H(|P|n)e^{-i\phi n\cdot J_{\rm f}}=H(|P|n),\ \ \ 
\phi\in\RR.$$
We  have 
\eq{dec}
\fffb=\bigoplus_{z\in {\Bbb Z}}\fffb^0(z),\ \ \ 
H(P)\cong H(|P|n)=\bigoplus_{z\in {\Bbb Z}} H(P,z)
\en 
where 
$\fffb^0(z)$ denotes the subspace spanned by eigenvectors associated 
with eigenvalue  $z\in{\Bbb Z}$ of $n\cdot J_{\rm f}$. 
It is shown that the ground state of $H(P)$ is unique for an arbitrary $e\in\RR$ 
in the case of  $P=0$, 
 and 
for a sufficiently small $|e|$ in the case of  $P\not=0$. See Section 3 and \cite{hisp2}. 
 
\bc{ss}
Let $\vp(R(n,\phi)k)=\vp(k)$ and 
polarization vectors be in Example \ref{lilo}. 
Assume that $H(|P|n)$ has a unique ground state $\gr(|P|n)$. 
Then $(n\cdot J_{\rm f})\gr(|P|n)=0$, 
i.e., $\gr(|P|n)\in \fffb^0(0)$ in the decomposition \kak{dec}.
\ec
\proof 
Since the ground state $\gr(|P|n)$ is unique,  
$\gr(|P|n)$ has to belong to  some $\fffb^0(z)$. 
Then it has to be $z=0$ since $H(|P|n,z)\cong H(|P|n,-z)$, $z\in{\Bbb Z}$,  by \cite{sasa}. 
\qed

\section{Functional integral representations}
In the quantum mechanics the functional integral representation of the semigroup 
$e^{-t h(a)}$ with 
$$\d h(a):=\half(-i\nabla-a)^2+V$$ on $\LR$ for  
real multiplication operators  
$a=(a_1,\cdots,a_d)$ and $V$ is  given  
through a stochastic integral. 
Let $(b(t))_{t\geq 0}=(b_1(t), \cdots, b_d(t))_{t\geq 0}$ be the $d $-dimensional Brownian motion starting at $0$ on a 
probability space $(W, {\cal B}, db)$. 
Set $X_s:=x+b(s)$, $x\in\BR$,  and $dX:=dx\otimes db$.
Then it is known that 
\eq{fki}
(f, e^{-th(a)} g)_\LR=
\int_{\BR\times W}  \ov{f(X_0)} g(X_t) e^{-i\int _0^t a (X_s)\circ db_s} dX,
\en 
where 
$\displaystyle \int _0^t a (X_s)\circ db(s) :=\sum_{\mu=1}^d 
\int _0^t a_\mu (X_s) db_{\mu}(s)+\half \int _0^t (\nabla\cdot  a) (X_s)ds.$
For the functional integral representation of the semigroup generated by the Pauli-Fierz Hamiltonian $H$  
we also need a stochastic integral but a Hilbert space-valued one.   
We quickly review a functional integral representation of $e^{-tH}$ in the next subsection.

\subsection{Functional integral representations for $e^{-tH}$}

\subsubsection{Gaussian random variables ${\cal A}_0,{\cal A}_1,{\cal A}_2$}
Let  ${\cal A}_0(f)$
be a Gaussian random process on a probability space $(Q_0, \Sigma_0, \mu_0)$
indexed by real $f=(f_1,...,f_d )\in \sso\LR$
with mean zero:
\eq{mean}
\int_{Q_0}{\cal A}_0(f)d\mu_0=0,
\en 
 and the covariance:
\eq{cov}
\int _{Q_0}{\cal A}_0(f){\cal A}_0(g) d\mu_0=q_0(f,g),
\en
where 
$$q_0(f,g):=
\half\sum_{\alpha,\beta=1}^d \int_{\BR}
 \dep_{\alpha\beta}(k) \ov{\hat f_\alpha(k)} \hat g_\beta(k) dk.
$$
The existence of probability space  
$(Q_0, \Sigma_0, \mu_0)$ and Gaussian random 
variable ${\cal A}_0(f)$
satisfying \kak{mean} and \kak{cov} are governed by
the Minlos theorem \cite[Theorem I.10]{si}.
In a similar way,  thanks to the Minlos theorem,  we can construct two other 
Gaussian random variables. 
Let 
${\cal A}_1(f)$ indexed by real $f\in \sso\LRT$  
 and  ${\cal A}_2(f)$ by real $f\in \sso\LRTT$ be 
 Gaussian random processes  on probability spaces  $(Q_1, \Sigma_1, \mu_1)$ and  $(Q_2, \Sigma_2, \mu_2)$, respectively, 
with mean zero and covariances given by 
\begin{eqnarray}
\label{ch1}
&&\int _{Q_1}{\cal A}_1(f){\cal A}_1(g)d \mu_1=q_1(f,g),\\
\label{ch11}
&&\int _{Q_2}{\cal A}_2(f){\cal A}_2(g)d \mu_2=q_2(f,g),
\end{eqnarray}
where 
\begin{eqnarray*}
&& q_1(f,g):=\half\sum_{\alpha,\beta=1}^d \int_{\RR^{d+1}}
 \dep_{\alpha\beta}(k) \ov{\hat f_\alpha (k,k_0)}
\hat g_\beta (k,k_0) dk dk_0,\\
&&
q_2(f,g):=
\half\sum_{\alpha,\beta=1}^d \int_{\RR^{d+1+1}}
 \dep_{\alpha\beta}(k) 
\ov{\hat f_\alpha (k,k_0,k_1)}
\hat g_\beta (k,k_0,k_1) dk dk_0 dk_1.
\end{eqnarray*}
Note that
${\cal A}_\#(f)$, $\#=0,1,2$, is real linear in $f$.
We extend it for  $f=f_{\rm R}+i f_{\rm I}$  with  
$f_{\rm R}=(f+\bar f)/2$ and $f_{\rm I}=(f-\bar f)/(2i)$ 
as ${\cal A}_\#(f)={\cal A}_\#(f_{\rm R})+i {\cal A}_\#(f_{\rm I})$.
The $n$-particle subspace $L_n^2(Q_\#)$ of $L^2(Q_\#)$ is defined by
$$L_n^2(Q_\#)=\ov{
\mbox{L.H.}\{:{\cal A}_{\#}(f_1) \cdots{\cal A}_{\#}(f_n):| f_j\in L^2
(\RR^{d +\#}),j=1,...,n\}}.$$
Here $:X:$ denotes the Wick product of $X$ \cite{rs2} defined recursively as  
\begin{eqnarray*}
&& :{\cal A}_{\#}(f):={\cal A}_{\#}(f),\\
&& :{\cal A}_{\#}(f){\cal A}_{\#}(f_1)\cdots{\cal A}_{\#}(f_n):=:{\cal A}_{\#}(f_1)\cdots{\cal A}_{\#}(f_n): \\
&& \hspace{1cm} -\sum_{j=1}^n q_{\#}(f,f_j) :{\cal A}_{\#}(f_1)\cdots\hat {{\cal A}_{\#}(f_j)}\cdots{\cal A}_{\#}(f_n):,
\end{eqnarray*}
where $\hat  Y$ denotes neglecting $Y$. 
The identity $L^2(Q_\#)=\oplus_{n=0}^\infty L_n^2(Q_\#)$ 
is known as the Wiener-Ito decomposition.

\subsubsection{Factorization of semigroups}
Let $T:L^2(\RR^{d +\#})\rightarrow L^2(\RR^{d +\#})$ 
be a linear contraction operator. 
Then the linear operator 
$\dgg T :L^2(Q_\#)\rightarrow L^2(Q_\#)$ is defined by
$$\dgg T 1=1,\ \ \
\dgg T
:{\cal A}_\#(f_1)\cdots {\cal A}_\#(f_n):=
:{\cal A}_\#(\K T_d  f_1)\cdots {\cal A}_\#(\K T_d  f_n):.$$
Since the linear hull of vectors of the form $:{\cal A}_\#(f_1)\cdots {\cal A}_\#(f_n):$ is dense in $L^2(Q_\#)$, 
we can extend $\dgg T$ to the contraction operator on $L^2(Q_\#)$. We denote its extension by the same symbol. 
In particular since
$\{\dgg{e^{it{ h}}}\}_{t\in\RR}$ with 
a self-adjoint operator ${h}$ on $\LR$
is a strongly continuous one-parameter unitary group, by the Stone theorem,
there exists a self-adjoint operator $\dggg {h}$ on  
$L^2(Q_\#)$ such that
$\dgg{e^{it{ h}}}=e^{it\dggg { h}}$, $t\in\RR$. 
We set
$N_\#:=\dggg 1$. 
Let $h$ be a multiplication operator in $\LR$. 
We define the families of isometries, 
\eq{xi}
j_s, \xi_t=\xi_t(h):
\LR\stackrel{j_s}{\longrightarrow}\LRT\stackrel{\xi_t}
{\longrightarrow}\LRTT,\ \ \ s,t\in\RR, 
\en
by 
\begin{eqnarray}
\label{ch2}
&&
\hat {j_s f}(k,k_0):=\frac{e^{-isk_0}}{{\sqrt\pi}}
\lk\frac{\omega(k)}{\omega(k)^2+|k_0|^2}\rk^\han \hat f(k),\ \ \ 
(k,k_0)\in\BR\times\RR,\\
&&\hat {\xi_t f}(k,k_0,k_1):=\frac{e^{-itk_1}}{{\sqrt\pi}}
\lk \frac{h(k)}{h(k)^2+|k_1|^2}\rk ^{\han} \hat f(k,k_0),
\ \ \ (k,k_0,k_1)\in \BR\times \RR\times \RR.\non
\end{eqnarray}
By a direct computation we can see that 
\begin{eqnarray}
\label{12}
&&
j_s^\ast j_t =e^{-|t-s|\omega(-i\nabla)}:\LR\rightarrow \LR\,\  \ \ s,t\in\RR, \\
&&
\label{122}
\xi_s^\ast \xi_t=e^{-|t-s|(h(-i\nabla)\otimes 1)}:\LRT\rightarrow \LRT,\ \ \ s,t\in\RR.
\end{eqnarray} 
Here
$\omega(-i\nabla)$ is  defined by
$\omega(-i\nabla)f =(\omega \hat f )^\vee$ 
and 
$h(-i\nabla)\otimes 1$ is an operator defined on $\LRT$ under the identification 
$\LRT\cong \LR\otimes L^2(\RR)$. 
Let us define the families  of operators $J_s$ and $\Xi_t=\Xi_t(h)$, $s,t\in\RR$; 
$$L^2(Q_0)\stackrel{J_s}{\longrightarrow} L^2(Q_1)\stackrel{\Xi_t}{\longrightarrow} L^2(Q_2)$$
by 
\begin{eqnarray*}
&&
J_s 1=1,\ \ \ J_s :
{\cal A}_0(f_1)\cdots {\cal A}_0(f_n): =
:{\cal A}_1(\K{j_s}_d  f_1)\cdots {\cal A}_1(\K{j_s}_d  f_n):,\ \ \ 
s\in\RR,\\
&&
\Xi_t 1=1,\ \ \ \Xi_t :
{\cal A}_1(f_1)\cdots {\cal A}_1(f_n): =
:{\cal A}_2(\K{\xi_t}_d  f_1)\cdots {\cal A}_2(\K{\xi_t}_d  f_n):,\ \ \ 
t\in\RR.
\end{eqnarray*}
Both $J_s$ and $\Xi_t$ can be extended to contraction operators in the similar manner as $\dgg T$.  
Those extensions are denoted by the same symbols. 
We have by \kak{12} and \kak{122} 
\begin{eqnarray}
\label{10}
&& J_s^\ast J_t=e^{-|t-s|d\Gamma_0{(\omega(-i\nabla))}}:L^2(Q_0)\rightarrow L^2(Q_0),\ \ \ s,t\in\RR, \\
&&\label {1000}
\Xi_s^\ast \Xi_t=e^{-|t-s|d\Gamma_1{(h(-i\nabla)\otimes 1)}}:L^2(Q_1)\rightarrow L^2(Q_1),\ \ \ s,t\in\RR.
\end{eqnarray}


\subsubsection{Functional integrals}
Define 
$${\cal A} _{\#,\mu}(f)=
{\cal A}_{\#}(\oplus_{\ell=1}^d  \delta_{\ell\mu} f),
\ \ \ f\in L^2(\RR^{d +\#}),\ \ \ \mu=1,...,d .$$
We set 
$$\d A_\mu(\hat f):=\frac{1}{\sqrt 2} 
\sum_{j=1}^{d-1}\int e_\mu (k,j) 
(\add(k,j){\hat f(k)} + a(k,j){\hat f(-k)} )dk$$ for $f\in\LR$, $\mu=1,...,d$. 
It is well known that $L^2(Q_0)$ is unitarily equivalent to
$\fffb$ with
$ 1\cong \Omega$, 
${\cal A}_{0,\mu}(f) \cong
A_\mu (\hat f)$ and 
$\DG {h(-i\nabla)}  \cong \dg h$.
In particular 
\eq{identification}
\DG{-i\nabla}\cong \pf,\quad \DG{\omega(-i\nabla)}\cong\hf
\en hold. 
Since  $\hhh\cong \int^\oplus_\BR\fffb dx$, 
 we can see that
\eq{u}
\hhh\cong \int^\oplus_\BR L^2(Q_0) dx,
\en
 i.e.,
$F\in\hhh$ can be regarded   as an $L^2(Q_0)$-valued $L^2$-function on $\BR$.
In what follows we use identification \kak{identification} and \kak{u} without notices.
Note that in the Fock representation the test function $\hat f$ of $A_\mu(\hat f)$  is taken in the 
momentum representation, but 
in the Schr\"{o}dinger representation,  $f$ of ${\cal A}_{0,\mu}(f)$   in the position representation.  
We can see that  
$$ H \cong \frac{1}{2}(-i\nabla\otimes 1-e{\cal A}_0^\la)^2+V\otimes 
1+1\otimes \hf .$$
Here ${\cal A}_0^\la:=({\cal A}_{0,1}^\la,...,{\cal A}_{0,d }^\la)$ with
$${\cal A}_{0,\mu}^\la:=\int_\BR^\oplus {\cal A}_{0,\mu}(\la(\cdot-x)) dx,\ 
\ \ \mu=1,...,d ,$$
and
$\la:=(\vp/\sqrt\omega)^\vee$. 
By the Feynman-Kac formula \kak{fki} with $a=(0,\cdots,0)$ and the fact $J_0^\ast J_t=e^{-t\hf}$ we can see that
$$( F, e^{-t H_0}G)_\hhh=
\int_{\BR\times W} e^{-\int_0^tV(X_s) ds}(J_0F(X_0), 
J_tG(X_t))_{L^2(Q_1)}dX.$$
Adding the minimal perturbation: $-i\nabla_\mu\otimes 1\rightarrow 
-i\nabla_\mu\otimes 1-e{\cal A}_0^\vp$, we can see in 
\cite{h4} the functional integral representation below.

\begin{proposition}
Let $F,G\in\hhh$. 
Then 
\eq{hi0}
(F, e^{-t H}G)_\hhh=
\int_{\BR\times W} e^{-\int_0^tV(X_s) ds}
(J_0F(X_0), e^{-ie{\cal A}_1({\cal K}_1^{[0,t]}(x))} J_tG(X_t))_{L^2(Q_1)}dX, 
\en 
where
$${\cal K}^{[0,t]}_1(x):=\oplus_{\mu=1}^d  \int_0^t j_s\la(\cdot-X_s) 
db_\mu(s)\in \oplus^d L^2(\RR^{d+1}),
$$
and 
\eq{hi00}
(F, e^{-tK} G)_\hhh=\int_{\BR\times W} e^{-\int_0^tV(X_s) ds}
(F(X_0), e^{-ie{\cal A}_0({\cal K}_0^{[0,t]}(x))}
G(X_t))_{L^2(Q_0)} dX,
\en 
where 
$${\cal K}_0^{[0,t]}(x)
=\oplus_{\mu=1}^d\int_0^t \tilde\varphi(\cdot-X_s) db_\mu(s)\in \oplus^d \LR.$$

\end{proposition}
Define two Gaussian random processes 
\begin{eqnarray*}
&&\{{\cal A}_{\mu, s}(f)\}_{s\in\RR,f\in\LR}   
\mbox{ on } (Q_1, \Sigma_1, \mu_1) 
\mbox{ by }
{\cal A}_{\mu, s}(f):={\cal A}_{1,\mu}(j_s f),\\
&&
\{{\cal A}_{\mu, s,t}(f)\}_{(s,t)\in\RR^2,f\in\LR} 
\mbox{ on } (Q_2, \Sigma_2, \mu_2) 
\mbox{ by }
{\cal A}_{\mu, s, t}(f):={\cal A}_{2,\mu}(\xi_s j_t f).
\end{eqnarray*}
By \kak{ch1}, \kak{ch11}, \kak{12} and \kak{122},  it is directly seen that 
\begin{eqnarray}
&&
\hspace{-1cm}
\label{vll}
\int _{Q_1}{\cal A}_{\alpha,s}(f) {\cal A}_{\beta, t}(g) d \mu_1=
\half \int_\BR  e^{-|s-t|\omega(k)}
 \dep_{\alpha\beta}(k) \ov{\hat f(k)}
\hat g (k) dk, \\
&&
\hspace{-1cm}
\label{vlll}
\int _{Q_2}{\cal A}_{\alpha,s,t}(f) {\cal A}_{\beta, s',t'}(g) d \mu_2=
\half \int_\BR e^{-|s-s'|h(k)} e^{-|t-t'|\omega(k)}
 \dep_{\alpha\beta}(k) \ov{\hat f(k)}
\hat g (k) dk. 
\end{eqnarray} 
Then the identity  
$\d {\cal A}_1({\cal K}_1^{[0,t]}(x)) =
\sum_{\mu=1}^d\int_0^t{\cal A}_{\mu, s}
(\tilde\varphi(\cdot-X_s)) db_\mu(s)$ follows.

\subsection{Functional integral representations for $e^{-t\fri(P)}$}
We shall construct the functional integral representation of  
$(\Psi, e^{-t\fri(P)}\Phi)_{\fff}$. 
\bl{cont}
Let $\Psi,\Phi\in\fffb$. Then $(\Psi, e^{-t\fri(P)}\Phi)_{\fff}$ and 
$(\Psi, e^{-tK_{\rm F}(P)}\Phi)_{\fff}$ 
are continuous in 
$P\in\BR$.
\el
\proof 
We prove the lemma for $f(P):=(\Psi, e^{-t\fri(P)}\Phi)_{\fff}$. 
That of $(\Psi, e^{-tK(P)}\Phi)_{\fff}$ is similar. 
We have 
\begin{eqnarray*}
&&
f(P)-f(P')=\int_0^t(e^{-(t-s)H(P)}\Psi, (H(P)-H(P'))e^{-sH(P')}\Phi)_\fffb ds \\
&&=
\frac{1}{2}\sum_{\mu=1}^d (P-P')_\mu 
\int_0^t(e^{-(t-s)H(P)}\Psi, (P+P'-2\pf-2e\av(0))_\mu e^{-sH(P')}\Phi)_\fffb ds.
\end{eqnarray*}
The integral on the right-hand side above is locally finite for $P$ and $P'$. Then 
it follows that $\lim_{P'\rightarrow P}f(P')=f(P)$ follows. 
\qed

For $\Psi\in L^2(Q_0)$, we set $\Psi_t:=J_t e^{-i\pf\cdot b(t)}\Psi$, $t\geq 0$. 
\bt{hiro1}
Let $\Psi,\Phi\in\fff$. Then 
\eq{kp}
(\Psi, e^{-t\fri(P)}\Phi)_{\fffb}=
\int_W 
(\Psi_0, e^{-ie{\cal A}_1(\kk{0,t})}
\Phi_t)_{L^2(Q_1)}\pb db,
\en 
where 
$\d \kk{0,t}:=\oplus_{\mu=1}^d  \int_0^t j_s\la(\cdot-b(s)) db_\mu(s), 
$
and 
\eq{kp2}
(\Psi, e^{-tK_{\rm F}(P)}\Phi)_{\fffb}=
\int_W 
(\Psi, e^{-ie{\cal A}_1(\kko{0,t})}
e^{-i\pf\cdot b(t)} \Phi)_{L^2(Q_0)} \pb db,
\en
where 
$\d \kko{0,t}:=\oplus_{\mu=1}^d  \int_0^t \la(\cdot-b(s)) db_\mu(s).$
\et
\proof
Set $F_s=\rho_s\otimes \Psi\in \LR \otimes \fffb_\infty$
and $G_{s'}=\rho_{s'}\otimes\Phi\in \LR\otimes \fffb_\infty$,
where $\rho_s$ is the heat kernel: 
\eq{heat}
\rho_s(x)=(2\pi s)^{-d /2}e^{-|x|^2/(2s)},\ \ \ s>0.
\en 
By the fact that $H=U\f\lk 
\int^\oplus_\BR \fri(P) dP\rk U $ and 
$Ue^{-i\xi\cdot \tot} U\f=\int_\BR^\oplus e^{-i\xi \cdot P} dP$, 
we have 
$$(F_s, e^{-tH} e^{-i\xi\cdot \tot}G_{s'})_\hhh
=\int_\BR dP ((U F_s)(P), e^{-t\fri(P)} e^{-i\xi\cdot P} 
(U G_{s'})(P))_\fffb,\ 
\ \ \xi\in \BR.$$
Here
$\d (U F_s)(P)=(2\pi)^{-d/2} \int_\BR
e^{-i x\cdot P} e^{ix\cdot \pf}  \rho_s(x) \Psi dx$. 
Note that
\eq{d1}
\lim_{s\rightarrow 0} (UF_s)(P)=\frac{1}{\sqrt{(2\pi)^d }}\Psi
\en
strongly in $\fffb$ for each $P\in\BR$.
Hence we have by the Lebesgue dominated convergence theorem,
\eq{esi2}
\lim_{s\rightarrow 0}
(F_s, e^{-tH} e^{-i\xi\cdot \tot}G_{s'})_\fff
=
\frac{1}{\sqrt{(2\pi)^d }}
\int_\BR dP (\Psi, e^{-t\fri(P)} e^{-i\xi\cdot P} (UG_{s'})(P))_\fffb
.
\en
On the other hand we see that by \kak{hi0}
\begin{eqnarray}
&&
\lim_{s\rightarrow 0}
(F_s, e^{-tH} e^{-i\xi\cdot \tot}G_{s'})_\hhh\nonumber \\
&&=
\lim_{s\rightarrow 0}
\int_{W\times \BR}
(J_0F_s(X_0), e^{-ie{\cal A}_1(\kkx)}
 J_t e^{-i\xi\cdot  \tot} G_{s'}(X_t))_{L^2(Q_1)}  dX\nonumber \\
&&
=\lim_{s\rightarrow 0}
\int_{W\times \BR}  \rho_s(X_0) \rho_{s'}(X_t-\xi)
(J_0\Psi, e^{-ie{\cal A}_1(\kkx)}  J_t e^{-i\xi\cdot \pf} 
\Phi)_{L^2(Q_1)}
 dX  \nonumber \\
&&
\label{esi3}
=\int_W    \rho_{s'}(b(t)-\xi)
(J_0\Psi, e^{-ie{\cal A}_1(\kkz)}
J_t e^{-i\xi\cdot \pf}  \Phi)_{L^2(Q_1)}db.
\end{eqnarray}
Here we used that
$e^{-i\xi\cdot \tilde{P}_T}(\rho(X_t)\otimes \Phi)
=\rho(X_t-\xi)\otimes e^{-i\xi\cdot\pf}\Phi$. 
The third equality  of \kak{esi3} is due to the Lebesgue dominated convergence 
theorem.
Then we obtained that from \kak{esi2} and \kak{esi3}
\begin{eqnarray}
&&
\frac{1}{\sqrt{(2\pi)^d }}
\int_\BR  e^{-i\xi\cdot P}
(\Psi, e^{-t\fri(P)} (UG_{s'})(P))_{\fff} dP\nonumber \\
&&\label{k1}
=
\int_W   \rho_{s'}(b(t)-\xi)
(J_0\Psi, e^{-ie{\cal A}_1(\kkz)} J_t e^{-i\xi\cdot \pf}
 \Phi)_{L^2(Q_1)} db.
\end{eqnarray}
Since 
$$\int_{\BR} \|e^{-t\fri(P)} UG_{s'}(P)\|_{\fff}^2 dP\leq 
\int_\BR\|UG_{s'}(P)\|^2_\fff dP=\|G_{s'}\|_\hhh^2<\infty,$$
we have 
$
(\Psi, e^{-t\fri(\cdot )}  (UG_{s'})(\cdot ))_\fffb\in\LR$ 
for $s'\not=0$. 
Then
taking the inverse Fourier transform
of the both-hand sides of \kak{k1} with respect to $P$, we have
\begin{eqnarray}
&&\hspace{-1cm} (\Psi, e^{-t\fri(P)} (UG_{s'})(P))_\fffb\nonumber \\
&&
\hspace{-1cm}
=
\frac{1}{\sqrt{(2\pi)^d }}
\int_\BR  d\xi e^{iP\cdot\xi}
\int_W  db   \rho_{s'}(b(t)-\xi)
(J_0\Psi, e^{-ie{\cal A}_1(\kkz)} J_t e^{-i\xi\cdot \pf}  
\Phi)_{L^2(Q_1)}\non\\
&&
\hspace{-1cm}
=\frac{1}{\sqrt{(2\pi)^d }}
\int_W  db
\int_\BR  d\xi e^{iP\cdot\xi}
\rho_{s'}(b(t)-\xi)
(J_0\Psi, e^{-ie{\cal A}_1(\kkz)} J_t e^{-i\xi\cdot \pf} 
\Phi)_{L^2(Q_1)}\non\\
&&\label{k2}
\end{eqnarray}
for almost every $P\in\BR$. The second equality of \kak{k2} is due to Fubini's 
lemma. 
The right-hand side of \kak{k2} 
is continuous in $P$, and 
the left-hand side is also continuous by  Lemma \ref{cont}. 
Then \kak{k2} is true for all $P\in \BR$.
Taking $s'\rightarrow 0$ on the both-hand  sides of \kak{k2}, we have by the 
Lebesgue dominated convergence theorem and \kak{d1},
$$
(\Psi, e^{-t\fri(P)} \Phi)_\fffb
=
\int_W
(J_0\Psi, e^{-ie{\cal A}_1(\kkz)} J_t \ppf \Phi)_{L^2(Q_1)}\pb db=\kak{kp}.
$$
Thus the theorem follows for $\Psi,\Phi\in\fffb_\infty$. 
Let $\Psi,\Phi\in\fffb$, and
$\Psi_n,\Phi_n\in\fffb_\infty$ such that
$\Psi_n\rightarrow\Psi$ and $\Phi_n\rightarrow \Phi$ strongly as 
$n\rightarrow \infty$.
Since
$$
|(J_0\Psi_n, e^{-ie{\cal A}_1(\kkz)}
J_t \ppf   \Phi_n)_{L^2(Q_1)}|\leq \|\Psi_n\|_\fffb
\|\Phi_n\|_\fffb \leq c$$
with some constant $c$ independent of $n$, we have by the Lebesgue dominated 
convergence theorem
\begin{eqnarray*}
&&
\limn
\int_W
(J_0\Psi_n, e^{-ie{\cal A}_1(\kkz)}
J_t \ppf  \Phi_n)_{L^2(Q_1)}\pb db\\
&&=\int_W 
(J_0\Psi, e^{-ie{\cal A}_1(\kkz)}
J_t \ppf  \Phi)_{L^2(Q_1)}\pb db,
\end{eqnarray*}
and it is immediate that
$\limn (\Psi_n, e^{-t\fri(P)} \Phi_n)_\fffb=(\Psi, e^{-t\fri(P)} 
\Phi)_\fffb$. 
Hence \kak{kp} is proved. 
\kak{kp2} is similarly proven through \kak{hi00} and the fact 
$\int_\BR^\oplus K_{\rm F}(P) dp\cong K$. 
\qed

\subsection{Applications}
Let 
$\d L^2_{\rm fin}(Q_\#) :=\bigcup_{N=0}^\infty [\oplus_{n=0}^N L^2_n(Q_\#)]$
and
$T$  a self-adjoint operator on $L^2(\RR^{d +\#})$.
Let us define operator $\Pi_{\#,\mu}(T f)$ 
on $L^2_{\rm fin}(Q_\#)$
by
\eq{p1}
\Pi_{\#,\mu}(T f):= i[d \Gamma_\# (T), {\cal A}_{\#,\mu}(f)],\ \ \ f\in D(T).
\en
In the case where $f$ is real-valued, $\Pi_{\#,\mu}(T f)$ 
is a
symmetric operator and 
$L^2_{\rm fin}(Q_\#)$ is the set of analytic 
vectors of
$\Pi_{\#,\mu}(f)$.
Then $L^2_{\rm fin}(Q_\#)$ is a core of
$\Pi_{\#,\mu}(f)$.
The self-adjoint extension of $\Pi_{\#,\mu}(f)$ with real $f$ is denoted by 
the
same symbol.

\subsubsection{Ergodic properties}
Let 
${\cal K}_+:=\{\Psi\in L^2(Q_0)| \Psi\geq 0\}$ be the positive cone and set 
 ${\cal K}_+^0:=\{\Psi\in{\cal K}_+|\Psi>0\}$. 
It is  well known \cite[Theorem I.12]{si} that 
$e^{i\pf \cdot v }{\cal K}_+\subset{\cal K}_+$ for $v\in \BR$.

\bp{pl}
For real $f\in L^2(\RR^{d +1})$, 
it follows that $J_0^\ast e^{i\Pi_{1,\mu}(f)} J_t[{\cal K}_+\setminus\{0\}]\subset{\cal K}_+^0$, i.e., 
$J_0^\ast e^{i\Pi_{1,\mu}(f)} J_t$ is positivity improving. 
\ep
\proof See \cite{fa1,fa2} for $f=0$ and \cite{h9} for $f\not=0$.
\qed
We define 
$$\t:=\exp\lk i\frac{\pi}{2}N\rk .$$
\bt{m2}
In the Schr\"odinger representation, 
 $\t  e^{-t\fri  (0)}\tf  $ is positivity improving. 
\et
\proof
Let $\Psi,\Phi\in{\cal K}_+\setminus\{0\}$. 
It is seen by the functional integral representation in Theorem \ref{hiro1} that
\begin{eqnarray}
(\Psi, \t  e^{-tH(0)} \tf   \Phi)_\fffb
&=& 
\int_W
(\Psi_0, e^{-ie{\Pi}_1(\kkz)}
\Phi_t)_{L^2(Q_1)} db\non\\
&=&\label{ir}
\int_W
(\Psi, J_0^\ast e^{-ie{\Pi}_1(\kkz)}
J_t \ppf \Phi)_{L^2(Q_0)} db.
\end{eqnarray}
Here we used the facts that   
$J_t \ppf e^{-i(\pi/2)N}=e^{-i(\pi/2)\tilde N}J_t\ppf$ and 
$$\d e^{i(\pi/2)\tilde N}e^{-ie{\cal A}_1(f)}e^{-i(\pi/2)\tilde N}=
e^{-ie\Pi_1(f)},$$
where $\tilde N=d\Gamma_1(1)$.  
By Proposition \ref{pl}, 
$J_0^\ast e^{-ie{\Pi}_1(\kkz)}
J_t \ppf$ is positivity improving for each $b\in W$. 
Namely the integrand in \kak{ir} is strictly positive for each $b\in W$. 
Hence the right-hand side of \kak{ir} is strictly positive, which implies that 
$ \t  e^{-tH(0)} \tf {\cal K}_+\setminus\{0\}\subset{\cal K}_+^0$.  
Thus the theorem follows.
\qed

\bc{c2}
The ground state $\gr(0)$ of $\fri (0)$ is unique up to multiple constants, if it exists, 
and it can be taken as $\t \gr(0)>0$ in the Schr\"odinger representation. 
\ec
\proof 
Theorem \ref{m2} implies that 
the ground state of $\t H(0)\tf$ is unique and that 
we can take a  strictly positive ground state,  
by an infinite dimensional version of the Perron-Frobenius theorem for a positivity improving operator. 
See \cite{glja}. Since $\t$ is unitary, the corollary follows. 
\qed

\bc{dia}{\bf [Two diamagnetic inequalities] }
It follows that 
\begin{eqnarray}
&&\hspace{-2cm}
 \label{dia1}
 |(\Psi,  e^{-t\fri(P)}  \Phi)_\fffb|
\leq 
(|\Psi|, e^{-t(\half\pf^2+\hf)} |\Phi|)_{L^2(Q_0)}, \\ 
&& \label{dia2}
\hspace{-2cm}
 |(\Psi, \t  e^{-t\fri(P)} \tf   \Phi)_\fffb|
\leq 
(|\Psi|, \t  e^{-t\fri(0)} \tf   |\Phi|)_{L^2(Q_0)}.
\end{eqnarray}
\ec
\proof 
When $L$ is positivity preserving,
it holds that $|L\Psi|\leq L|\Psi|$. 
We have 
$$
|(\Psi,  e^{-tH(P)}  \Phi)_\fffb|
\leq 
\int_W
(J_0|\Psi|, 
J_t \ppf |\Phi|)_{L^2(Q_1)}  db=(|\Psi|, e^{-t(\half\pf^2+\hf)} |\Phi|)_{L^2(Q_0)}
$$
where we used that $b(t)$ is Gaussian with $\int  |b_\mu(t)|^2 db=1/2$. 
Thus \kak{dia1} follows. 
We have
\eq{dia3}
(\Psi, \t  e^{-tH(P)} \tf   \Phi)_\fffb
=\int_W
(\Psi_0, e^{-ie{\Pi}_1(\kkz)}
\Phi_t)_{L^2(Q_1)} \pb db.
\en
Then  it follows that
\begin{eqnarray*}
|(\Psi, \t  e^{-t\fri(P)} \tf   \Phi)_\fffb|
&\leq &
 \int_W
(|\Psi|_0, e^{-ie{\Pi}_1(\kkz)}
|\Phi|_t)_{L^2(Q_1)} db\\
&=& 
(|\Psi|, \t  e^{-t\fri(0)} \tf   |\Phi|)_{L^2(Q_0)}. 
\end{eqnarray*}
Hence \kak{dia2} follows. 
\qed
Let 
$E(P,  e^2)=\inf\s(\fri(P))$. 
\bc{m3} 
(1)  
$0=E(0,0)\leq E(0,e^2)\leq E(P,e^2)$, 
(2) 
Assume that the ground state $\gr(0)$ of $H(0)$ exists for  
$e\in [0,e_0)$ with some $e_0>0$.  
Then $E(0,e^2)$ is concave, continuous and monotonously increasing function on $e^2$,   
(3) $E(0,e^2)\leq \is (H)$. 
\ec 
\proof 
\kak{dia2} implies 
$|(\Psi, \t e^{-tH(P)}\tf\Psi)_\fffb|\leq e^{-tE(0,e^2)}\|\Psi\|^2_\fffb$. 
Since $\t$ is unitary, (1) follows. 
Let $\gr(0)$ be the ground state of $H(0)$. 
Thus by Corollary \ref{c2}, $\t \gr(0)>0$, and hence $(1,\gr(0))_{L^2(Q_0)}\not=0$ by $\tf 1=1$. 
Thus
\begin{eqnarray*}
&&E(0,e^2)=
\lim_{t\rightarrow \infty} -\frac{1}{t}\log (\Omega, 
e^{-tH(0)}\Omega)_\fffb=
\lim_{t\rightarrow \infty} -\frac{1}{t}\log \int_W
(1, e^{-ie{\cal A}(\kk{0,t})}1 )_{L^2(Q_1)}   db\\
&&=\lim_{t\rightarrow \infty} -\frac{1}{t}\log
\int_W e^{-\frac{e^2}{2}q_0(\kkz, \kkz) }   
db.
\end{eqnarray*}
Since 
$e^{-\frac{e^2}{2}q_0(\kkz, \kkz) }$ is log convex on $e^2$,  $E(0,e^2)$ is concave. 
Then $E(0,e^2)$ is continuous on $(0,e_0)$. 
Since $E(0,e^2)$ is also continuous at $e^2=0$ by the fact that $H(0)$ converges as $e^2\rightarrow 0$ 
in the uniform resolvent sense, $E(0,e^2)$ is continuous on $[0,e_0)$. Then 
$E(0, e^2)$ can be expressed as $E(0, e^2)=\int_0^{e^2}\phi(t) dt$ 
with some positive function $\phi$. Thus 
$E(0, e^2)$ is monotonously increasing on $e^2$. 
Then (2) is obtained. 
We have 
$$(F, (1\otimes \t) e^{-tH} (1\otimes \tf) G)_\hhh=
\int_\BR dP(F(P), \t  e^{-t\fri(P)} \tf   G(P)))_\fffb.
$$ 
Then 
by \kak{dia2} it is seen that 
$$|(F, (1\otimes \t) e^{-tH} (1\otimes \tf) F)_\hhh|\leq 
e^{-tE(0,e^2)} \int_\BR dP\|F(P)\|_\fffb^2=e^{-tE(0,e^2)}\|F\|_\hhh^2.
$$ 
Thus (3) follows. 
\qed
\begin{remark}
(1) 
The uniqueness of the ground state of $H(P)$ is shown in \cite{hisp2} for 
a sufficiently small $|e|$. 
The result in Corollary \ref{c2} is valid for arbitrary values of 
coupling constants but $P=0$. 
(2) In \cite {lms}, a weaker statement $\t  e^{-tH(0)} \tf   {\cal K}_+\subset {\cal K}_+$ is shown, 
and Corollary~\ref{m3} (1) is also obtained.  
\end{remark}

\subsubsection{Invariant domains and essential self-adjointness of $K(P)$}
\label{332}
\bl{re1}
Assume that $\omega^{3/2}\vp\in\LR$. 
Then 
\eq{s1}
e^{-tK_{\rm F}(P)} [D(\pf^2)\cap D(\hf)]\subset
D(\pf^2)\cap D(\hf).
\en 
\el
\proof  
We have for $f\in\sso \LRT$,
\begin{eqnarray*}
e^{ie{\cal A}_0(f)} \hf   e^{-ie{\cal A}_0(f)}
&=& 
\hf -ie[\hf, {\cal A}_0(f)]
 +
\half (-ie)^2[[\hf,{\cal A}_0(f)], {\cal A}_0(f)]\\
&=& \hf-e\Pi_0(\K {\omega}_d f)-e^2 q_0(\K {\omega}_d f, f).
\end{eqnarray*}
From the Burkholder type inequality \cite{h11}: for $\mu=1,...,d$,
\eq{B}
\int_W
db  \left\|
\omega^{n/2}
\int_0^t \la(\cdot-X_s)db_\mu(s)\right\|_\LR^{2}
\leq
\frac{(2m)!}{2^m}t^m\|\omega^{(n-1)/2}\vp\|_\LR^{2}, 
\en
and  
$\| \Pi_0(f)\Phi\|_{L^2(Q)}
\leq c \sum_{\alpha=1}^d 
( \|f_\alpha/\sqrt\omega \|_\LR+
\|f_\alpha\|_\LR)
\|(\hf+1)^\han\Phi\|_{L^2(Q)}$ by 
\kak{aaa} with some constant $c$, 
it follows that for $\Psi\in D(\hf)$,
\eq{bu1}
\int_W \| e^{ie{\cal A}_0(\ZK)}
\hf 
e^{-ie{\cal A}_0(\ZK)}J_t \Psi\|_{L^2(Q_0)} ^2 db 
\leq
C \|(\hf+1)\Psi\|_\fffb^2
\en 
with some constant $C$.
Thus
we see that
by means of functional integral representations \kak{kp2},  
\begin{eqnarray}
&& |(\hf \Psi, e^{-tK_{\rm F}(P)}\Phi)_\fffb|\non \\
&& \leq
\int_W
\|e^{ie{\cal A}_0(\ZK)} \Psi\|_{L^2(Q_0)}
\|
e^{ie{\cal A}_0(\ZK)} \hf
e^{-ie{\cal A}_0(\ZK)}
\ppf \Phi\|_{L^2(Q_0)} db\non \\
&&\label{a}
\leq C' \|\Psi\|_\fffb \|(\hf+1)\Phi\|_\fffb
\end{eqnarray}
with some constant $C'$.
We can also see that 
\begin{eqnarray*}
&& 
e^{ie{\cal A}_0(f)} 
\pf ^2   
e^{-ie{\cal A}_0(f)}\\
&& =
\sum_{\mu=1}^d 
\lk
\pf_\mu -ie[\pf_\mu, {\cal A}_0(f)]+
\half (-ie)^2[[\pf_\mu, {\cal A}_0(f)], {\cal A}_0(f)]\rk^2\\
&& =
\sum_{\mu=1}^d 
\lkk 
\pf_\mu^2-e \pf_\mu  \Pi_0(\K{-i\nabla_\mu}_d f)-e\Pi_0(\K{-i\nabla_\mu}_d  f)
\pf_\mu 
+e^2\Pi_0(\K{-i\nabla_\mu}_d f)^2 \right.\\
&&\left. 
-e^2q_0 (\K{-i\nabla_\mu}_d  f,f) \pf_\mu
+e^3 \Pi_0(\K{-i\nabla_\mu}_d  f)q_0 (\K{-i\nabla_\mu}_d f,f)+
e^4 
q_0 (\K{-i\nabla_\mu}_d f, f)^2\rkk.
\end{eqnarray*}
We have for $\Psi\in D(\pf^2)\cap D(\hf)$
\begin{eqnarray}
\label{Ba}
&&\|\pf_\mu \Pi_0(\K{-i\nabla_\mu}_d f)\Psi\|\leq c_1\|(\pf^2+\hf +1)\Psi\|,\\
&&\|\Pi_0(\K{-i\nabla_\mu}_d f)\pf_\mu\Psi\|\leq c_2\|(\pf^2+\hf+1)\Psi\|,\\
&&\|\Pi(\K{-i\nabla_\mu}_d f)^2\Psi\|\leq c_3\|(\hf+1)\Psi\|,\\
&&\|q_0(\K{-i\nabla_\mu}_d f,f)\pf\Psi\|\leq c_4\|\pf \Psi\|,\\
&&\|\Pi_0(\K{-i\nabla_\mu}_d f) q_0(\K{-i\nabla_\mu}_d f,f)\Psi\|\leq c_5\|\hf^\han \Psi\|,\\
\label{Bak}
&& \|q_0(\K{-i\nabla_\mu}_d f,f)^2\Psi\|\leq c_6\|\Psi\|,
\end{eqnarray}
where we used $\omega^{3/2}\vp\in\LR$ in \kak{Ba}. 
Thus, together with Burkholder type inequality \kak{B}, the integration of $c_1,...,c_6$  
in \kak{Ba} - \kak{Bak} with $f$ replaced by ${\cal K}_0^{[0,t]}(0)$ over $db$ 
is suppressed from upper and 
\begin{eqnarray}
&& |(\pf^2 \Psi, e^{-tK_{\rm F}(P)}\Phi)_\fffb|\non \\
&& \leq
\int_W
\|e^{ie{\cal A}_0(\ZK)} \Psi\|_{L^2(Q_0)}
\|
e^{ie{\cal A}_0(\ZK)} \pf^2
e^{-ie{\cal A}_0(\ZK)}
 \ppf \Phi\|_{L^2(Q_0)} db\non \\
&&\label{aa}
\leq C' \|\Psi\|_\fffb \|(\hf+\pf^2+1)\Phi\|_\fffb
\end{eqnarray}
with some constant  $C'$. Here we used that $\|\hf \ppf \Psi\|=\|\hf \Psi\|$. 
By \kak{a} and \kak{aa} the lemma is obtained. 
\qed
{\bf Proof of Theorem \ref{poo} (2)}\\
\proof 
By Lemma \ref{re1} we see that $K_{\rm F}(P)$ is essentially self-adjoint on $D(\pf^2)\cap D(\hf)$ by 
\cite[Theorem X.49]{rs2}. Since $K_{\rm F}(P)=K(P)$ on $D(\hf)\cap D(\pf^2)$, the desired result is obtained. 
\qed

\bl{n}
Let  $\Phi\in D((N+1)^\alpha)$ with $\alpha\in{\Bbb N}$. 
Then $\ehtp \Phi\in D(N^\alpha)$ with the inequality 
$\|N^\alpha \ehtp \Phi\|_\fffb 
\leq 
C\|(N+1)^\alpha\Phi\|_\fffb$,  
where $C$ is a constant independent of~$P$. 
\el
\proof 
Let $\Psi,\Phi\in D(N^\alpha)$. Set ${\cal A}={\cal A}_1(\kkz)$ and  $\tilde N=d\Gamma_1(1)$ for simplicity. 
Then 
\eq{mn}
(N^\alpha \Psi, \ehtp \Phi)=
\int_W(\tilde N^\alpha \Psi_0, 
e^{-ie{\cal A}} \Phi_t)_{L^2(Q_1)} \pb db.
\en 
Since 
\begin{eqnarray}
e^{ie{\cal A}} \tilde N^\alpha  e^{-ie{\cal A}} \Phi
&=& \lk \tilde N-ie[\tilde N, {\cal A}]+
\frac{1}{2!}(-ie)^2[[\tilde N,{\cal A}],{\cal A}]\rk^\alpha\Phi\non \\
&=&\label{nn}
\lk \tilde N-e\Pi_1(\kkz)-e^2q_1(\kkz , \kkz)\rk^\alpha \Phi.
\end{eqnarray}
The right-hand side of \kak{nn} is suppressed as 
$\|\mbox{r.h.s.}\kak{nn}\|_{L^2(Q_1)} \leq c \|(N+1)^\alpha\Phi\|_\fffb
$
by 
the Burkholder type inequality 
\eq{BB}
\int_W
db  \left\|
\int_0^t j_s\la(\cdot-X_s)db_\mu(s)\right\|_\LRT^{2}
\leq
\frac{(2m)!}{2^m}t^m\|\vp/\sqrt\omega \|_\LR^{2}.
\en
Then  
\begin{eqnarray*}
&&\hspace{-1cm}
|(N^\alpha \Psi, \ehtp \Phi)_\fffb|
\leq 
\int_W |(\Psi_0, 
\tilde N^\alpha  e^{-ie{\cal A}} 
 \Phi_t) _{L^2(Q_1)} | 
db\\
&&\hspace{-1cm}
 \leq \int_W \| e^{ie{\cal A}} \Psi_0\|_{L^2(Q_1)}
\|e^{ie{\cal A}} \tilde N^\alpha  e^{-ie{\cal A}} 
 \Phi_t\|_{L^2(Q_1)}db
\leq  c\|\Psi\|_\fffb  \|(N+1)^\alpha\Phi\|_\fffb.
\end{eqnarray*}
Hence the lemma follows. 
\qed

\section{The $n$ point Euclidean Green functions}
In this section we extend functional integral representations derived 
in the previous section to the $n$ point Euclidean Green functions. 
We fix $0=s_0\leq s_1\leq \cdots \leq s_{m-1}\leq s_m=s$ and 
$0=t_0\leq t_1\leq \cdots \leq t_{m-1}\leq t_m=t$. 
For notational simplicity we define for objects (operators or vectors)  $T_j$, $j=1,...,n$, 
$$\prod_{j=1}^n T_j:=T_1T_2\cdots T_n.$$
We introduce the  set $\fffi$ of bounded operators on $\fff$ by 
$$\fffi:=\{\Phi(A(f_1),\cdots,A(f_n))| \Phi \in L^\infty(\RR^n), f_j\in \oplus^d \LR, j=1,...,n,n\geq 0\}.$$
We identify bounded multiplication operator  
$\Phi(A(f_1),\cdots,A(f_n))$ on $\fff$ and 
bounded multiplication operator $\Phi({\cal A}_0(f_1),\cdots,{\cal A}_0(f_n))$ on $L^2(Q_0)$.

\subsection{In the case of $H$}
\bt{main3}
Let  $K=1\otimes d\Gamma(h)$ with a  multiplication operator $h$ in $\LR$. 
Let
$F_j=f_j\otimes\Phi_j\in L^\infty(\BR)\otimes \fffi$, $j=1,...,m-1$, 
with 
$\Phi_j=\Phi_j(A(f^j_1),\cdots, A(f_{n_j}^j))$, and $F_0,F_m\in\hhh$.  
Then 
\begin{eqnarray}
&&
(F_0, \prod_{j=1}^m e^{-(s_j-s_{j-1}) K} 
e^{-(t_j-t_{j-1}) H}F_j)_{\hhh} \non\\
&&\label{F}
=
\int_{\BR\times W}  
e^{-\int_0^tV(X_s) ds}
( \hat F_0(X_0),
e^{-ie{\cal A}_2({\cal K}_2(x))}  
\prod_{j=1}^{m} \hat F_j(X_{t_j})   
)  _{L^2(Q_2)}dX,
\end{eqnarray}
where $\hat F_j(x):=\Xi_{s_j} J_{t_j} F_j(x)=
f_j(x) \Phi_j({\cal A}_2(\xi_{s_j}j_{t_j} f_1^j),\cdots, {\cal A}_2(\xi_{s_j}j_{t_j} f_{n_j}^j))$  
and  
$$\d {\cal K}_2(x):=\oplus_{\mu=1}^d \sum_{j=1}^m 
 \int_{t_{j-1}}^{t_j} \xi_{s_j} j_s\la(\cdot-X_s) db_\mu(s).
 $$
 \et
\proof 
Set $\d K_0=\half(-i\nabla\otimes 1-e{\cal A}_0^{\tilde{\vp}})^2$ and assume that 
$V\in C_0^\infty(\BR)$ in a moment. 
By the Trotter-Kato product formula \cite{mk}, 
we have 
$e^{-tH}=s-\lim_{n\rightarrow \infty} \TPP t .$
Set $a_n=t_n-t_{n-1}$ and $b_n=s_n-s_{n-1}$, $n=1,...,m$,  for notational convenience.  
Thus 
\begin{eqnarray}
&&\hspace{-0.7cm}
\mbox{l.h.s. }\kak{F}=\limn (F, \T {b_1} \TPP{a_1} F_1 \T {b_2} \TPP{a_2}\cdots\non \\
&&\label{FFF}\hspace{3cm}
\cdots F_{m-1} \T {b_m} \TPP {a_m} G)_{L^2(Q_0)}.
\end{eqnarray}
Define 
$Q_s:\hhh\rightarrow \hhh$ by 
\begin{eqnarray*}
&& (Q_0 F)(x):=F(x),\\
&& (Q_sF)(x):=\int_{\BR} p_s(|x-y|) e^{-i(e/2)\sum_{\mu=1}^d {\cal A}_{0,\mu}(\la(\cdot-x)+\la(\cdot-y))\cdot(x_\mu-y_\mu)}F(y) dy,\ \ \ s\not=0.
\end{eqnarray*}
Here 
$p_s(x)$ is the heat kernel given in \kak{heat}.
Then it is established in \cite{h4} that 
\eq{FF}
s-\limn (Q_{t/2^n})^{2^n}=e^{-tK_0}.
\en 
Let 
$E_s:=J_s J_s^\ast$ and define 
$
Q_{[a,b]}:=\mbox{L.H.} \{F\in L^2(Q_1)| F\in {\rm Ran} E_s, s\in [a,b]\}$. 
We denote the smallest $\sigma$ field generated by $Q_{[a,b]}$ by $\Sigma_{[a,b]}$. 
Let $a\leq b\leq c\leq d$ and 
assume that 
$\Psi$ is measurable with respect to $\Sigma_{[a,b]}$ and 
$\Phi$ with respect to $\Sigma_{[c,d]}$. Then it is known as Markov property \cite{si} of $E_s$ on $L^2(Q_1)$ 
that 
\eq{mar}
(\Psi, E_s\Phi)_{L^2(Q_1)}=(\Psi, \Phi)_{L^2(Q_1)}
\en 
for $b\leq s\leq c$. 
We note that for $F=f\otimes \Phi({\cal A}_0(f_1),\cdots, {\cal A}_0(f_n))
\in L^\infty(\BR)\otimes\fffi$, the identity 
\eq{most}
 J_s F J_s^\ast=  (J_s F) E_s =E_s (J_s F) E_s
\en 
holds as an operator, 
where  $J_sF$ 
 on the right-hand side of \kak{most}
is 
$$J_s F=f\otimes \Phi({\cal A}_1(j_{s} f_1),\cdots, {\cal A}_1(j_{s} f_n)).$$
In particular it follows that 
\eq{mostt}
J_s e^{-i{\cal A}_{0,\mu}(f)} J_s^\ast= E_se^{-i{\cal A}_{1,\mu}(j_s f)} E_s
\en 
as an operator. 
Substituting \kak{FF},  
$e^{-|s-t|\hf} =J_s^\ast J_t$ and  
$ J_s e^{-td\Gamma_1(h)}= \TT t J_s$, 
where 
$\tilde K:=d\Gamma_1(h\otimes 1)$,   
into \kak{FFF}, 
we can obtain that 
\begin{eqnarray*}
&&\hspace{-0.5cm}\mbox{l.h.s }\kak{F}=\int_{\BR\times W} dX e^{-\int_0^t V(X_s) ds} 
(J_0 F_0(X_0), \TT {b_1} e^{-ie{\cal A}_1(K_{t_0}^{t_1})}
(J_{t_1} F_1(X_{t_1})) \TT {b_2} \cdots\\
&&\hspace{2cm}
\cdots (J_{t_{m-1}} F_{m-1}(X_{t_m})) \TT {b_m}  e^{-ie{\cal A}_1(K_{t_m}^{t_{m-1}})} J_{t_m} 
F_m(X_{t_m}))_{L^2(Q_1)},
\end{eqnarray*}
where we used 
\kak{most}, \kak{mostt}  and   the Markov property \kak{mar} of $E_s$,  
and set simply $\d K_u^v={\cal K}_1^{[u,v]}(x)=\oplus_{\mu=1}^d \int_u^v j_s\la(\cdot-X_s) db_\mu(s)$. 
Factorizing $\TT {b_j}$ as $\Xi_{s_j}^\ast \Xi_{s_{j-1}}$ and 
using the Markov property of $\Xi_s\Xi_s^\ast$ on $L^2(Q_2)$ again, we have 
\begin{eqnarray*}
&&\mbox{l.h.s. }\kak{F}=\int _{\BR\times W} dX e^{-\int_0^t V(X_s) ds} 
(\Xi_0 J_0 F_0(X_0), e^{-ie{\cal A}_2(\xi_{s_1} K_{t_0}^{t_1})}e^{-ie{\cal A}_2(\xi_{s_2} K_{t_1}^{t_2})}\cdots \\
&&\hspace{2cm} \cdots e^{-ie{\cal A}_2(\xi_{s_m} K_{t_{m-1}}^{t_m})} 
\prod_{j=1}^{m} (\Xi_{s_j} J_{t_j} F_j(X_{t_j})) 
)_{L^2(Q_2)}=\mbox{r.h.s. } \kak{F}.
\end{eqnarray*}
Hence the theorem follows for $V\in C_0^\infty(\BR)$. By a simple limiting argument 
on $V$, we can get the theorem.  
\qed
\begin{remark}\label{extended}
By the proof of Theorem \ref{m2}, 
we can see that  
$F_1,...,F_{m-1}$ in Theorem~\ref{m2} can be extended for more general bounded multiplication operators
such as the form  
$F(x)=e^{-ix\cdot\pf} f(x) \otimes \Psi(A(f_1),\cdots,A(f_n))e^{ix\cdot \pf} =
f(x) \otimes \Psi(A(e^{-ikx} f_1),\cdots,A(e^{-ikx}f_n))$. 
This facts will be used in the next subsection. 
\end{remark}

\subsection{In the case of $\fri(P)$}

\bt{pmain}
Let $K=d\Gamma(h)$ with a multiplication operator $h$ in $\LR$.  
We assume that $\Phi_0,\Phi_m\in\fff$ 
and $\Phi_j\in \fffi$ for $j=1,...,m-1$ with 
$\Phi_j=\Phi_j(A(f^j_1),\cdots,A(f_{n_j}^j))$.
Then for  $P_0, \cdots,P_{m-1}\in\BR$, 
\begin{eqnarray}
&& 
(\Phi_0, \prod_{j=1}^m 
\xxx j  {j -1}  \Phi_j)_\fffb \nonumber \\
&&\label{left}
=\int_W  
(\hat \Phi_0, e^{-ie{\cal A}_2({\cal K}_2(0))} 
\prod_{j=1}^{m}  \hat \Phi_j
)_{L^2(Q_2)}e^{+i\sum_{j=1}^m (b(t_j)-b(t_{j-1})) P_{j-1}} db,
\end{eqnarray}
where 
$$\hat\Phi_j:=\Xi_{s_j}J_{t_j}e^{-i\pf \cdot b(t_j)}\Phi_j=
\Phi_j({\cal A}_2(\xi_{s_j}j_{t_j}f^j_1(\cdot-b(t_j))), \cdots, 
{\cal A}_2(\xi_{s_j}j_{t_j}f^j_{n_j}(\cdot-b(t_j))))$$  
and 
$$\d {\cal K}_2(0):=\oplus_{\mu=1}^d\sum_{j=1}^m
\int_{t_{j-1}}^{t_j}\xi_{s_j}j_s\tilde\vp(\cdot-b(s))db_\mu(s).
$$ 
In particular, in the case of $P_0=\cdots =P_{m-1}=P$, it follows that 
$$
(\Phi_0, \prod_{j=1}^m \xxxx j {j-1} \Phi_j)_\fffb
=\int_W   (\hat \Phi_0, e^{-ie{\cal A}_2({\cal K}_2(0))} 
\prod_{j=1}^{m}  \hat \Phi_j
)_{L^2(Q_2)}e^{iP\cdot b(t)} db.
$$
\et
\proof 
Let $\xi_0,\xi_1,\cdots,\xi_{m-1}\in\BR$ and $l_0,l_1,\cdots,l_{m-1}>0$. 
Set  
$F_j(x)=\rho_{l_j}(x)\Phi_j(x)$, where 
$\Phi_j(x)=e^{-ix\cdot\pf}\Phi_je^{ix\cdot\pf}=
\Phi_j({\cal A}_0(f_1^j(\cdot-x)),\cdots,{\cal A}_0(f_{n_j}^j(\cdots-x)))$, 
and 
$\rho_s$ is the heat kernel given in \kak{heat}. 
Then 
\eq{46}
UF_j\Psi= 
(\hat\rho_{l_j}\Phi_j)\ast(U\Psi)=
\left[
\int_{\BR}\hat\rho_{l_j}(\cdot-y) (U\Psi)(y) dy \right]
\Phi_j
\en 
follows, where $U$ is given in \kak{esi12} and $\hat\rho$ the Fourier transform of $\rho$.  
For notational convenience we set,  for $j=1,...,m$,  
$$
O_j(P_{j-1}):=\xxx j {j-1},\quad 
O_j:=e^{-(s_j-s_{j-1})1\otimes K} e^{-(t_j-t_{j-1})H}.
$$
Then the left-hand side of \kak{left} can be presented as 
$
\can$ 
We note that  
$[O_j(P), e^{-i\eta\cdot \tot}]=0$, $P,\eta\in\BR$. 
Set $\tilde F_j(P)=\hat\rho_{l_j}(P)\Phi_j$. 
We see that by \kak{46} 
\begin{eqnarray}
&& (F_0, \yyy 0 O_1  F_1 \yyy 1 O_2  F_2\cdots \yyy {m-1} O_m F_m)_\fffb\non \\
&&= (UF_0, U \yyy 0 O_1  F_1
 \yyy 1 O_2  F_2\cdots \yyy {m-1} O_m F_m)_\fffb\non \\
&&=(2\pi)^{-d/2}\int_\BR dP_0(\tilde F_0(P_0), O_1(P_0)[
\tilde F_1\ast(U O_2\yyy 1\cdots F_m)](P_0))_\fffb e^{-i\xi_0\cdot P_0}\non \\
&&=[(2\pi)^{-d/2}]^2 \int_\BR dP_0\int_\BR dP_1 
(\tilde F_0(P_0), O_1(P_0) \tilde F_1(P_0-P_1) O_2(P_1)\non \\
&&\hspace{3cm} \times [\tilde F_2 
\ast(U O_3\yyy 2\cdots F_m)](P_1))_\fffb\yyyy 0 \yyyy 1\non \\
&&\hspace{6cm}\vdots \non \\
&&\hspace{6cm}\vdots \non \\
&&=[(2\pi)^{-d/2}]^m \int_{(\BR)^m} dP
\bar{\hat{\rho}}_{l_0}(P_0) \hat\rho_{l_m}(P_{m-1})
\prod_{j=1}^{m-1}\hat{\rho}_{l_j}(P_{j-1}-P_j)
\non \\
&&\hspace{7cm}\times 
\label{sankaku} \can e^{-i\xi\cdot P},
\end{eqnarray}
where $P=(P_0,\cdots, P_{m-1})\in(\BR)^m$ and 
$\xi=(\xi_0,\cdots,\xi_{m-1})\in(\BR)^m$. 
On the other hand, we can see that 
\begin{eqnarray}
&&\hspace{-0.5cm}
 (F_0, \yyy 0  O_1 F_1   \yyy 1 O_2 F_2\cdots \yyy{m-1}O_m F_m)_\fffb\non \\
&& \hspace{-0.5cm}= (F_0, \yyy 0  O_1 F_1e^{+i\xi_0\tot} e^{-i(\xi_0+\xi_1)\cdot 
\tot}
 O_2 F_2e^{+i(\xi_0+\xi_1)\cdot \tot} \cdots 
e^{-i(\xi_0+\cdots+\xi_{m-1})\cdot 
\tot}O_m F_m)_\fffb\non \\
&&\hspace{-0.5cm}=(\dot{F}_0(0),  O_1 \dot{F}_1(\xi_0)   O_2 \dot{F}_2(\xi_0+\xi_1)\cdots 
O_m \dot{F}_m(\xi_0+\cdots+\xi_{m-1}))_\fffb \non \\
&&\hspace{-0.5cm}=\int_{\BR\times W} \rho_{l_0}(X_{t_0})\rho_{l_1}(X_{t_1}-\xi_0)\cdots 
\rho_{l_m}(X_{t_m}-\xi_0-\cdots-\xi_{m-1})   \non \\
&& 
\label{sankaku2}
\hspace{5cm}
\times (\hat \Phi_0(X_0),  e^{-ie{\cal A}_2({\cal K}_2(x))} 
 \prod_{j=1}^{m}
\hat \Phi_j(X_{t_j}))_{L^2(Q_2)}dX,
\end{eqnarray}
where 
$\dot{F}_j(\xi)=\rho_{l_j}(x-\xi)\Phi_j(x)$, 
$\hat\Phi_j(x)=
\Phi_j({\cal A}_2(\xi_{s_j}j_{t_j} f^j_1(\cdot-x)),\cdots, {\cal A}_2( \xi_{s_j}j_{t_j}f^j_{n_j}(\cdot-x)))$, 
 and the third identity above is due to 
Theorem \ref{main3} and Remark \ref{extended}. 
Hence by \kak{sankaku} and \kak{sankaku2} we obtain that 
\begin{eqnarray}
&&\hspace{-0.5cm}[(2\pi)^{-d/2}]^m \int _{(\BR)^m} e^{-i\xi\cdot P} 
\bar{\hat{\rho}}_{l_0}(P_0) \hat\rho_{l_m}(P_{m-1})
\prod_{j=1}^{m-1}\hat{\rho}_{l_j}(P_{j-1}-P_j)
\can\non \\
&&\hspace{-0.5cm}
=\int_{\BR\times W} \rho_{l_0}(X_{t_0})\lk\prod_{j=1}^{m-1}
\rho_{l_j}(X_{t_j}-\sum_{i=0}^{j-1}\xi_i) \rk
(\hat \Phi_0(X_0),  e^{-ie{\cal A}_2({\cal K}_2(x))} 
 \prod_{j=1}^{m}
\hat \Phi_j(X_{t_j}))_{L^2(Q_2)} dX. \non\\
&& \label{poi}
\end{eqnarray}
Since 
$\can$  is bounded with respect to $P\in(\BR)^m$, 
the integrand of the left-hand side of \kak{poi} 
satisfies that 
$$\bar{\hat{\rho}}_{l_0}(P_0) \hat\rho_{l_m}(P_{m-1})
\prod_{j=1}^{m-1}\hat{\rho}_{l_j}(P_{j-1}-P_j)\can 
\in L^2((\BR)^m, dP).
$$
By this, we can take 
the inverse Fourier transform of the both-hand sides of \kak{poi} in the $L^2$-sense.
Then we obtain that 
\begin{eqnarray}
&&\bar{\hat{\rho}}_{l_0}(P_0) \hat\rho_{l_m}(P_{m-1})
\prod_{j=1}^{m-1}\hat{\rho}_{l_j}(P_{j-1}-P_j)
\can 
\non \\
&&=[(2\pi)^{-d/2}]^m\int_{(\BR)^m}  d\xi e^{+i\xi \cdot P} 
\int_{\BR\times W} \rho_{l_0}(X_{t_0})\lk\prod_{j=1}^{m-1}
\rho_{l_j}(X_{t_j}-\sum_{i=0}^{j-1}\xi_i) \rk\non\non \\
&&\hspace{4cm}
\times 
\label{POL}(\hat \Phi_0(X_0),  e^{-ie{\cal A}_2({\cal K}_2(x))} 
\prod_{j=1}^{m}
\hat \Phi_j(X_{t_j}))_{L^2(Q_2)} dX
\end{eqnarray}
for almost every $P\in (\BR)^m$. Since the both-hand sides of \kak{POL} are continuous in 
$P\in(\BR)^m$ by Lemma \ref{cont}.
Thus \kak{POL} is true for all $P\in(\BR)^m$. 
Take the limit, ${l_m\rightarrow 0}\cdots {l_0\rightarrow 0}$, on the both-hand sides of \kak{POL}. 
On the left-hand side of \kak{POL} 
we have 
\begin{eqnarray}
&&\lim_{l_m\rightarrow 0}\cdots \lim_{l_0\rightarrow 0} 
\bar{\hat{\rho}}_{l_0}(P_0) \hat\rho_{l_m}(P_{m-1})
\prod_{j=1}^{m-1}\hat{\rho}_{l_j}(P_{j-1}-P_j)\can 
 \non \\
&&\label{f22}=[(2\pi)^{-d/2}]^m 
\can. 
\end{eqnarray}
We note that $\hat\Phi_j(X_{t_j})\lceil_{x=0}=\hat\Phi_j$ by the definition of $\hat\Phi_j$.  
Then we see that,  
on the right-hand side of \kak{POL} without coefficient $[(2\pi)^{-d/2}]^m$,  
\begin{eqnarray}
&&\lim_{l_m\rightarrow 0}\cdots \lim_{l_0\rightarrow 0} 
\int_{(\BR)^m} d\xi 
 e^{+i\xi \cdot P} 
\int_{\BR\times W} dX\rho_{l_0}(X_{t_0})\lk\prod_{j=1}^{m}
\rho_{l_j}(X_{t_j}-\sum_{i=0}^{j-1}\xi_i) \rk\non\\
&&\hspace{7cm}
\times 
(\hat \Phi_0(X_0),  e^{-ie{\cal A}_2({\cal K}_2(x))} 
\prod_{j=1}^{m}
\hat \Phi_j(X_{t_j}))\non \\
&&=
\lim_{l_m\rightarrow 0}\cdots \lim_{l_1\rightarrow 0} 
\int_{(\BR)^m}  d\xi 
e^{+i\xi \cdot P} 
\int_{W} db \lk\prod_{j=1}^{m}
\rho_{l_j}(b({t_j})-\sum_{i=0}^{j-1}\xi_i) \rk
\non\\
&&\hspace{7cm}
\times 
(\hat \Phi_0,  e^{-ie{\cal A}_2({\cal K}_2(0))} 
\prod_{j=1}^{m}
\hat \Phi_j))\non  \\ 
&&=
\lim_{l_m\rightarrow 0}\cdots \lim_{l_2\rightarrow 0} 
\int_{(\BR)^{m-1}} d\xi_1\cdots d\xi_{m-1}   
e^{+i(b(t_1)-b(t_0))P_0+i\sum_{j=1}^{m-1}\xi_j \cdot P_j} \non \\
&&\hspace{2cm}
\times 
\int_{W} db \lk\prod_{j=2}^{m}
\rho_{l_j}(b({t_j})-b(t_1)-\sum_{i=1}^{j-1}\xi_i) \rk
(\hat \Phi_0,  e^{-ie{\cal A}_2({\cal K}_2(0))} 
\prod_{j=1}^{m}
\hat \Phi_j) \non \\ 
&&=
\lim_{l_m\rightarrow 0}\cdots \lim_{l_3\rightarrow 0} 
\int_{(\BR)^{m-2}}d\xi_2\cdots d\xi_{m-1}    
e^{+i(b(t_1)-b(t_0))P_0+i(b(t_2)-b(t_1))P_1+i\sum_{j=2}^{m-1}\xi_j \cdot P_j} \non \\
&&\hspace{2cm}
\times 
\int_{W} db \lk\prod_{j=3}^{m}
\rho_{l_j}(b({t_j})-b(t_2)-\sum_{i=2}^{j-1}\xi_i) \rk\non
(\hat \Phi_0,  e^{-ie{\cal A}_2({\cal K}_2(0))} 
 \prod_{j=1}^{m}
\hat \Phi_j) \non \\
&&\hspace{5cm}\vdots \non \\
&&\hspace{5cm}\vdots \non \\&&\label{f23}
=
\int_W db  e^{+i\sum_{j=1}^m (b(t_j)-b(t_{j-1})) P_{j-1}} 
(\hat \Phi_0, e^{-ie{\cal A}_2({\cal K}_2(0))} 
\prod_{j=1}^{m}  \hat \Phi_j
).
\end{eqnarray}
From \kak{f22} and \kak{f23} the theorem follows. 
\qed

\bc{A}
Let $\Psi,\Phi\in\ffff$. Then 
\begin{eqnarray}
&&\hspace{-1cm}
(\Psi, e^{-t_1\fri(P)}A(f_1) e^{-(t_2-t_1)H(P)} A(f_2)\cdots A(f_{n-1})e^{-(t_n-t_{n-1})H(P)}\Phi)_\fff\non \\
&&\hspace{-1cm}
\label{AAAA}=\int _W  (\Psi_0, e^{-ie{\cal A}_1({\cal K}_1^{[0,t]}(0))}
\lk 
\prod_{j=1}^{n-1} 
{\cal A}_1(j_{t_j}f(\cdot- b(t_j)))\rk  \Phi_t)_{L^2(Q_1)}\pb db.
\end{eqnarray}
\ec
\proof 
We show an outline of a proof. 
We note that the left-hand side of \kak{AAAA} is well defined by Lemma \ref{n}.
First we can see by Theorem \ref{pmain} that 
\begin{eqnarray}
&&\hspace{-1cm}
(\Psi, e^{-t_1\fri(P)}e^{i s_1A(f_1)} e^{-(t_2-t_1)H(P)} e^{is_2A(f_2)}
\cdots e^{is_{n-1}A(f_{n-1})}
e^{-(t_n-t_{n-1})H(P)}\Phi)_\fff\non \\
&&\hspace{-1cm}
\label{AAAAA}=\int _W 
(\Psi_0, e^{-ie{\cal A}_1({\cal K}_1(0)^{[0,t]})}
\lk 
\prod_{j=1}^{n-1} e^{is_j 
{\cal A}_1(j_{t_j}f(\cdot- b(t_j)))}\rk 
 \Phi_t)_{L^2(Q_1)}\pb db.
\end{eqnarray}
By  Lemma \ref{n}, $e^{-t_1 H(P)}\Psi\in C^\infty (N)$. Then 
$e^{is_1A(f_1)} e^{-t_1 H(P)}\Psi$ is strongly differentiable at $s_1=0$ with 
$$\frac{d}{ds_1} e^{is_1A(f_1)} e^{-t_1 H(P)}\Psi\lceil_{s_1=0}=iA(f_1)e^{-t_1 H(P)}\Psi, $$
and 
$e^{is_2A(f_1)} e^{-(t_2-t_1)  H(P)}A(f_1)e^{-t_1 H(P)}\Psi$
is also differentiable at $s_2=0$ with 
$$\frac{d}{ds_2} e^{is_2A(f_2)} e^{-(t_2-t_1)  H(P)}A(f_1)e^{-t_1 H(P)}\Psi
= A(f_2) e^{-(t_2-t_1)  H(P)}
A(f_1)e^{-t_1 H(P)}\Psi.$$
Repeating this procedure on the left-hand side of \kak{AAAAA}, we can get the left-hand side of \kak{AAAA}.
It is also seen that the right-hand side of \kak{AAAAA} is 
also differentiable at $(s_1,...,s_{n-1})=(0,...,0)$ with the right-hand side of \kak{AAAA} as a result. 
Thus the corollary follows. 
\qed

\subsection{Applications}
We shall show some application of Theorem \ref{pmain}, by which 
we can construct a sequence of measures on $W$  converging to 
$(\gr(P), T\gr(P))_\fffb$ for some bounded operator $T$.  
In particular $T=e^{-\beta N}$ and $T=e^{-iA(f)}$ are taken as examples. 
In \cite{hisp2} 
it is proved that $H(P)$ has a unique ground state $\gr(P)$ and 
 $(\gr(P), \Omega)_\fffb\not=0$ for a sufficiently small $e$.

\bc{ex}
We suppose that  $H(P)$ has the  unique ground state $\gr(P)$ and 
it satifies $(\gr(P),\Omega)_{\fff}\not=0$. Then for $\beta>0$, 
$$(\gr(P), e^{-\beta N}\gr(P))=\lim_{t\rightarrow\infty}
 \int _W e^{(e^2/2)(1-e^{-\beta}) D(t)} e^{iP\cdot b(2t)}d\mu_{2t},$$
where 
$D(t):=q_1(\kk{0,t},\kk{t,2t})$
and $\mu_{2t}$ is a measure on $W$ given by 
$$d\mu_{2t}:=\frac{1}{Z}
{e^{-(e^2/2)q_1(\kk{0,2t},\kk{0,2t})}  }db$$
with the normalizing constant $Z$ such that 
$\int_W  e^{iP\cdot b(2t)} d\mu_{2t}=1$. 
\ec
\proof 
We define the family of isometries $\xi_s=\xi_s(1)$, $s\in\RR$,  by \kak{xi}. 
By Theorem~\ref{pmain} we have 
\begin{eqnarray*}
(e^{-t\fri(P)} \Omega,e^{-\beta N}e^{-t\fri(P)}\Omega)_\fffb
&=&
\int_W db e^{iP\cdot b(2t)}(1, 
e^{-ie{\cal A}_2(\xi_0\kk{0,t}+\xi_\beta\kk{t,2t})} 1)_{L^2(Q_2)}\\
&=& 
\int_W db e^{iP\cdot b(2t)}e^{-(e^2/2)
q_2(\xi_0\kk{0,t}+\xi_\beta\kk{t,2t})} .
\end{eqnarray*}
Noticing that $q_2(\xi_s f,  \xi_tg)=e^{-|s-t|}q_1(f,g)$, we have 
$$
q_2(\xi_0\kk{0,t}+\xi_\beta\kk{t,2t})
=
q_1(\kk{0,2t},\kk{0,2t})-(1-e^{-\beta})q_1(\kk{0,t},\kk{t,2t}).
$$
Then 
\eq{beta}
 \frac{(e^{-t\fri(P)} \Omega,e^{-\beta N}e^{-t\fri(P)}\Omega)}
{
(e^{-t\fri(P)} \Omega, e^{-t\fri(P)}\Omega)}=
\int _W e^{(e^2/2)(1-e^{-\beta}) D(t)} e^{iP\cdot b(2t)}d\mu_{2t}.
\en
The corollary follows from   the fact 
$$s-\lim_{t\rightarrow \infty} 
\frac{e^{-tH(P)}\Omega}{\|e^{-tH(P)}\Omega\|_\fff}
=\frac{(\gr(P),\Omega)_\fffb}{|(\gr(P),\Omega)_\fff|}
\gr(P)$$ and \kak{beta}. 
\qed

\begin{remark}\label{exx}
It is informally written as 
\begin{eqnarray*}
&& 
q_1(\kk{S,T},\kk{S', T'})\\
&&
=\half
\sum_{\alpha,\beta=1}^d \int_S^T db_\alpha(s)\int_{S'}^{T'}db_\beta(s')
\int_\BR\dep_{\alpha\beta}(k) \frac{|\vp(k)|^2}{\omega(k)} 
e^{-|s-s'|\omega(k)}e^{-ik(b(s)-b(s'))}dk.
\end{eqnarray*}
There are some discussions on double stochastic integrals mentioned above in 
\cite{sp5}.
\end{remark}

\bc{AA}
Assume the same assumptions as in Corollary \ref{ex}. 
Then 
\eq{pe}
(\gr(P), e^{-iA(f)}\gr(P))_\fffb=\limt \int_W
e^{-e q_1(\ZZ, f^t)-\half q_0(f,f)}
e^{iP\cdot b(2t)} d\mu_{2t},
\en
where 
$f^t:=\oplus_{\mu=1}^d j_t f_\alpha(\cdot-b(t))$.
\ec
\proof 
We have by Theorem \ref{pmain}
\begin{eqnarray*}
(\gr(P), e^{-iA(f)}\gr(P))_\fffb
&=&
\limt  \frac{
(e^{-tH(P)}\Omega, e^{-iA(f)}
e^{-tH(P)}\Omega)_\fffb}
{(e^{-tH(P)}\Omega, e^{-tH(P)}\Omega)_\fffb}\\
&=&
\limt \frac{1}{Z}
\int _W db 
e^{iP\cdot b(2t)}
(1, e^{
-i(e{\cal A}_1(\ZZ)+
{\cal A}_{1}(j_tf))} 1)_{L^2(Q_1)}\\
&=&
\limt \frac{1}{Z}
\int _W db e^{iP\cdot b(2t)}
e^{-\half q_1(e\ZZ+ f^t)}. 
\end{eqnarray*}
Note that 
$\d q_1(f^t,f^t)=
q_0(f,f)$. Then the corollary follows. 
\qed
\begin{remark}
$q_1(\ZZ,f^t)$ is informally  given by 
$$
q_1(\ZZ, f^t)=\half\sum_{\alpha,\beta=1}^ d \int_0^{2t}db_\alpha(s)
\int_\BR\delta^\perp_{\alpha\beta}(k) 
\frac{\ov{\vp(k)}}
{\sqrt{\omega(k)}}
\hat f_\beta(k) e^{ik\cdot(b(s)-b(t))}e^{-|s-t|\omega(k)}dk.
$$
\end{remark}

{\bf Acknowledgment}
I thank Sasaki for a useful comment on a rotation invariance. 
This work is partially done in 
"Open Quantum System'' 
held in Ervin Schr\"odinger Institute (ESI) on March of 2005, and 
"Feynman-Kac Formulas and their Applications"  
in the Wolfgang Pauli Institute (WPI) of Universit\"at Vienna on March of 2006.
The author thanks these kind hospitalities.
This work is financially supported by 
Grant-in-Aid  for Science Research (C) 17540181 from JSPS.

\end{document}